\documentclass[12pt]{article}

\usepackage{graphicx,latexsym,amsmath,amssymb,float,appendix,subfig}
\newcommand{\texorpdfstring}[2]{#1}
\graphicspath{ {Figures_and_Images/}}
\DeclareGraphicsExtensions{.pdf,.png,.jpg}

\newtheorem{thm}{\bf{Theorem}}[section]
\newtheorem{lem}[thm]{\bf{Lemma}}

\newtheorem{df}[thm]{\bf{Definition}}

\newtheorem{prop}[thm]{\bf{Proposition}}

\newtheorem{ex}[thm]{\bf{Example}}

\newenvironment{tightenumerate}{\begin{list}{--~}{
  \topsep=0.3ex \itemsep=0.3ex \labelsep=1em \parsep=0em
  \listparindent=0em \itemindent=0em
  \settowidth{\labelwidth}{--~} \leftmargin=3.5em
}}{\end{list}}

\newenvironment{param}{\medskip \begin{tightenumerate}}{\end{tightenumerate}\medskip}

\newenvironment{alg}{{\medskip \noindent \bf{Algorithm: }}
    \begin{tabbing}}{\end{tabbing}\medskip}

\newcommand{\Proof}{\noindent {\bf Proof:} }

\newcommand{\A}{A}
\newcommand{\B}{B}

\newcommand{\x}{x}
\newcommand{\y}{y}
\newcommand{\z}{z}
\newcommand{\h}{\omega_h}
\newcommand{\vv}{\omega_v}
\newcommand{\V}{V}
\newcommand{\E}{E}
\newcommand{\vertex}{v}

\newcommand{\current}{current}
\newcommand{\neighbour}{neighbour}

\newcommand{\hp}{\h^\prime}
\newcommand{\vp}{\vv^\prime}
\newcommand{\Vp}{\V^\prime}
\newcommand{\Ep}{\E^\prime}
\newcommand{\twoDO}{\text{2D-$\omega$}}
\newcommand{\threeDO}{\text{3D-$\omega$}}
\newcommand{\maxDiff}{\delta C_{max}}
\newcommand{\minDiff}{\delta A_{min}}

\newcommand{\activeM}{active_{heap}}
\newcommand{\sVertex}{v_s}
\newcommand{\dVertex}{v_d}
\newcommand{\mxy}{m_{x,y}}
\newcommand{\mxiyi}{m_{x_i,y_i}}
\newcommand{\T}{T}

\newcommand{\kk}{k} 
\newcommand{\kkk}{\texorpdfstring{$k$}{k}}%
\newcommand{\ka}{\kappa}
\newcommand{\kb}{q}
\newcommand{\ww}{w}
\newcommand{\RR}{R}
\newcommand{\Hm}{H_{max}}
\newcommand{\Hi}{H_i}

\newcommand{\SD}{s \text{--} d}

\newcommand{\Path}[1]{p_{#1}}
\newcommand{\Pathp}{p}
\newcommand{\Pathq}{q}
\newcommand{\penaltyWidth}{w_p}
\newcommand{\penaltyMax}{w_{max}}
\newcommand{\areaDiff}{\delta A}
\newcommand{\PathLength}{L}
\newcommand{\bigO}{O}

\begin{document}

\title{Multiple-Path Selection for new Highway Alignments using Discrete Algorithms
\thanks{\copyright 2015. This manuscript version is made available under the CC-BY-NC-ND 4.0 license http://creativecommons.org/licenses/by-nc-nd/4.0/}
}
\author{Yasha Pushak\footnote{ASC 303, Computer Science, Unit 5 Arts \& Sciences, UBC Okanagan, 3333 University Way, Kelowna BC V1V 1V7},
Warren Hare\footnote{ASC 353, Mathematics, Unit 5 Arts \& Sciences, UBC Okanagan, 3333 University Way, Kelowna BC V1V 1V7},
Yves Lucet\footnote{yves.lucet@ubc.ca, ASC 350, Computer Science, Unit 5 Arts \& Sciences, UBC Okanagan, 3333 University Way, Kelowna BC V1V 1V7}}
\maketitle

\begin{abstract}
	This paper addresses the problem of finding multiple near-optimal, spatially-dissimilar paths that can be considered as alternatives in the decision making process, for finding optimal corridors in which to construct a new road. We further consider combinations of techniques for reducing the costs associated with the computation and increasing the accuracy of the cost formulation. Numerical results for five algorithms to solve the dissimilar multipath problem show that a ``bidirectional approach'' yields the fastest running times and the most robust algorithm. Further modifications of the algorithms to reduce the running time were tested and it is shown that running time can be reduced by an average of 56\% without compromising the quality of the results.
\end{abstract}
	
\section{Introduction}
\label{sec:Introduction}
	
	In preliminary road design, selecting the best path for a new road is traditionally a long and political process.  A variety of factors can require that road engineers design multiple alternative paths to be considered. Previous research has developed methods of modeling the road's costs as well as computing the optimal alignment~\cite{TURNER-78,TRIETSCH-87}. While this software provides useful insight for road engineers, it does not satisfy the need to find multiple road path options for review. There is a need for an algorithm that can efficiently compute multiple near-optimal, but distinctly different, paths that can serve as alternatives to the cheapest path.
	
	This paper is concerned with the first step of the road design process: corridor selection. An initial path can then be refined, optimizing the horizontal alignment within the corridor, using techniques from Mondal et al.~\cite{MONDAL-14}, Hare et al.~\cite{HARE-11,HARE-14,HARE-14b}, or citations therein.
	
	To locate the initial corridor we model the terrain as a three dimensional grid of points to construct a spatial graph. Adapted forms of Dijkstra's Algorithm are then used to select the set of spatially-dissimilar corridors, while minimizing an approximation of the earthwork and pavement costs. Additional techniques to reduce the running time have also been developed. These include an adapted form of the A* algorithm and two methods of imposing height restrictions, to avoid spending time searching for unrealistic roads that are far from the ground.

\subsection{Road Design}\label{sec:Road Design}
	
			Previous work on road alignment selection has developed a number of discrete models. The simplest and most common is a regular grid~\cite{TURNER-78}, where the center of each grid cell is given a vertex and then edges are defined by the vertices of the eight adjacent grid cells. This model restricts possible road trajectories to having eight directions. Another method, presented by Trietsch~\cite{TRIETSCH-87,TRIETSCH-87a}, used a honey-comb grid, which allows for angles in multiples of $30\,^{\circ}$. In many previous discrete models the space between each adjacent vertex was as small as 200 m and up to 2~km~\cite{TRIETSCH-87,TRIETSCH-87a,TURNER-78}, which does not allow for detailed and accurate assessment of costs in a given region of a cell. Many models also formulated the construction costs independently of the road's direction~\cite{TURNER-78}.
			
			On the horizontal alignment problem, Huber and Church~\cite{HUBER-85} took an in-depth look at minimizing the errors in the cost evaluation associated with path planning problems, such as road design, by increasing the number of possible directions of movement. Later, Lee et al.~\cite{LEE-09} applied a neighbourhood search technique to approximate horizontal alignments with a piecewise-linear curve. Once they selected a horizontal alignment they refined it to be a smooth path meeting road standards such as curvature restraints. Easa and Mehmood~\cite{EASA-08} used collision frequency data for various road types to improve the safety of horizontal alignments. Kang et al.~\cite{KANG-07} improved an existing Horizontal Alignment Optimization (HAO) model by restricting the search space with feasible gates.
			
			Kang~\cite{KANG-08} used genetic algorithms to choose new highway segments that intersect with an existing highway network. A bi-level approach was developed to first select candidate paths with intermediate solutions to the genetic algorithm and then evaluate them for traffic flow optimization~\cite{KANG-10, KANG-12}.
			
			Jha~\cite{JHA-03} approached the road design problem with genetic algorithms. Further work was done to integrate GIS to include the costs of land boundaries, environmental impact, topography, travel time, and noise and air pollution of highways near cities~\cite{JHA-04,JHA-06}. Jha et al.~\cite{JHA-06a} reviewed cost formulation as well as common road design optimization algorithms.	Continuing this work Yang et al.~\cite{YANG-14} adapted the genetic algorithms to find alternate routes. Bosurgi et al. considered environmental constraints using particle swarm optimization~\cite{BOSURGI-13}. They also added new types of curves and provided a genetic algorithm approach to optimize the parameters~\cite{BOSURGI-14}. Shafahi and Bagherian proposed a customized particle swarm optimization algorithm~\cite{SHAFAHI-13} while another 3D highway alignment model solved by evolutionary algorithms was considered by Jong and Schonfeld~\cite{JONG-03}. It is also worth mentioning Jong's Phd Thesis~\cite{JONG-98}.
			
			User interface designs have been developed and proposed for the road design problem. Church et al.~\cite{CHURCH-92} designed one using a multi-objective model to allow finding both optimal and alternative paths.	
			
			Many of the previous road design optimization methods are focused on finding a single path or corridor. In practice, road design is a political process in which it is impossible to determine and evaluate all of the environmental and political cost factors ahead of time. Rather than using a multi-objective model with what will surely be an incomplete set of costs and constraints, this paper focuses on finding $\kk$ spatially dissimilar paths of nearly equal cost that can be reviewed by a board of engineers, politicians, and/or environmentalists for additional costs and impacts.

		\subsection{\kkk-shortest-path Algorithms}
		
			Extensive research has been done on the problem of finding the $k$-shortest-paths in a network~\cite{BRANDER-95,eppstein97,HOFFMAN-59, LAWLER-72,YEN-71}. In general there are two different approaches: deletion algorithms~\cite{BRANDER-95,LAWLER-72,YEN-71} and deviation algorithms~\cite{eppstein97,HOFFMAN-59}.
			
			Deletion algorithms propose using a conventional path-finding algorithm, such as Dijkstra's, to find the optimal path. The edges from the optimal path are successively deleted from the graph and the path-finding algorithm is re-run to generate a set of secondary paths. The cheapest among them is selected and the process iterates.
	
			Deviation algorithms use the information generated by a shortest-path tree to the destination to exploit the frequent locality of the $k$-shortest-paths~\cite{eppstein97,HOFFMAN-59}. They begin with the cheapest path and then search for the deviation that offers the smallest increase in cost. While this property of the $k$-shortest-paths is what allows them to achieve the most competitive time complexity, it also offers insight as to why it, and indeed any of the conventional $k$-shortest-path algorithms, are not well-suited for adaptation to the spatially-dissimilar multipath problem.
	
			Both families of algorithms perform poorly when applied to the multipath road design problem, which corresponds to a $k$-shortest-path problem applied to a dense spatial graph\footnote{We call a dense spatial graph, a spatial graph with vertices having degree 26 except on the boundary.} with the additional complication of finding spatially dissimilar paths.
			
\begin{figure}[ht]
				\includegraphics[scale=0.8]{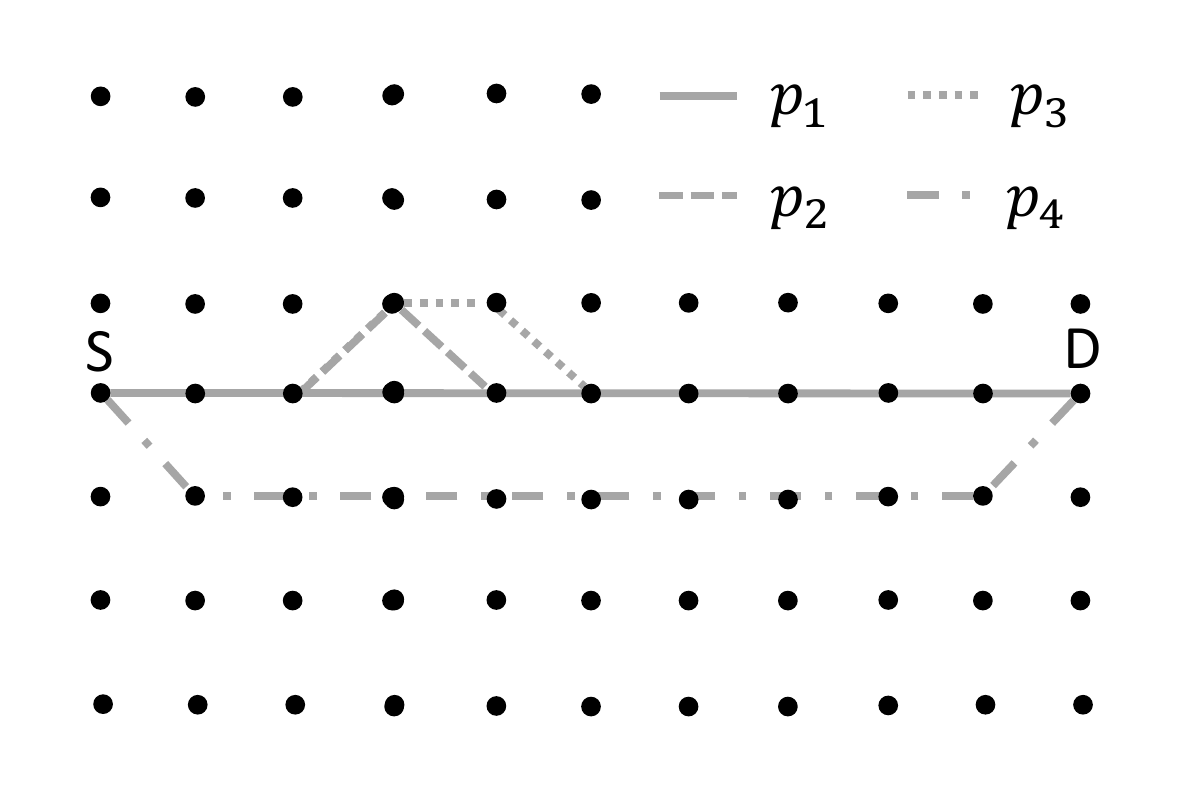}
				\centering
				\caption{A uniform cost grid and four possible paths illustrating the expected behaviour of a $k$-shortest-path algorithm. The cheapest path is $\Path1$, and possible second, third, and fourth cheapest paths are $\Path2$, $\Path3$, and $\Path4$.}
				\label{fig:bad-KSP}
			\end{figure}

			As an academic example consider the behaviour of a $k$-shortest-path algorithm on a uniform cost grid (something that loosely approximates a very flat prairie). The globally optimal path would be the straight line $\Path1$ seen in Figure \ref{fig:bad-KSP}. Paths $\Path2$, $\Path3$, and $\Path4$ would all be equally priced and so any of them may be the second path found depending on how the algorithm breaks ties. Of the three options presented $\Path{4}$ would rank as the most desirable since it is the most different from $\Path{1}$. However, $\Path{4}$ is still nearly identical to $\Path{1}$ when using a refined grid and so none of these paths should be considered acceptable as candidate alternate paths to $\Path{1}$.
		
			The simplest way to adapt these algorithms to the multipath road-design problem is to iteratively generate paths until we have found $k$ paths that satisfy our spatial dissimilarity criterion. Since early iterations of this approach would be spatially similar paths, this method would have to be iterated quite extensively before good alternative routes are found. As such, this approach performs poorly in practice.

			The dissimilar path problem is one often proposed in the context of transporting hazardous materials across highway networks~\cite{DELL-05,JOHNSON-92,KANG-13}. The goal of the dissimilar multipath problem is to find a set of spatially dissimilar paths between a source and a destination. A variety of dissimilarity indices have been used to approach this problem. For two given paths many of them define a relationship between the shared length of the paths and the non-shared length of the paths~\cite{AKGUN-00,KANG-13,KUBY-97}. This method is well-suited to the hazardous material (HazMat) objective function which is based on the value of risk rather than the transportation costs. It yields local minima that are spatially diverse. Dissimilar path algorithms are better suited to finding alternative paths for road design than standard $\kk$-shortest-path algorithms. However, in the context of road design this method will find paths such as $\Path4$ seen in Figure \ref{fig:bad-KSP}. These paths will share a minimal amount of length, but will be spatially very similar, due to the dense nature of the road design grid. Indeed, minimizing the shared length of paths is best suited to existing highway networks where alternate paths consist of different highway options.
			
			Dell'Olmo et al.~\cite{DELL-05}, also in the context of HazMat transportation, introduces an interesting modification to previous dissimilarity indices by defining a buffer of area around each of the paths and calculating an index determined by the area of overlap. This method is particularly useful in HazMat transportation where the buffer area can be computed as the expected area that would be affected by an accident.
			
			Another metric used by Marti et al.~\cite{MARTI-14,MARTI-09} is a possible option that could have provided meaningful candidate alternative paths for road design. The distance metric used by Marti computes the average of the shortest distance from each vertex on path $\Path1$ to $\Path2$ and vice versa. The average distances of each path are normalized by the lengths of the respective paths and then the average of the two paths is taken as the dissimilarity index.
			
			While both of the previous two methods have the potential to provide meaningful candidate alternative paths for road design, we opted to use another dissimilarity criterion, similar to one proposed by Lombard and Church~\cite{LOMBARD-93}. Their metric looks at the area difference between two paths using a method analogous to computing the absolute difference of the area under two curves. This method was chosen out of the three since it has equal potential to providing significantly different paths, while requiring the least computational expense.
			
			Most of the dissimilar path algorithms in the literature use a dissimilarity criterion related to the shared length of the paths and many only compare the alternate paths to the single cheapest path rather than to each other. This paper uses the area difference between two paths as a metric to ensure that the paths are spatially different. This type of metric is required for the dense graph used for the road design model, since it ensures that small deviations as seen in Figure \ref{fig:bad-KSP} are not accepted for alternate paths. This paper also measures the area between each path to ensure that no two alternate paths are spatially similar.
				
		\subsection{Dissimilar Path Generation}
		\label{sec:Dissimilar Path Generation}
			
			Several types of algorithms have been used to generate paths for the various dissimilarity criteria. A standard method among them is the Iterative Penalty Method~\cite{AKGUN-00,JOHNSON-92,ROUPHAIL-96}. Turner~\cite{TURNER-78} applied a similar method to the multipath road design problem by increasing the weights of the edges of the original path. We adapt this idea for the road design problem with spatially dense graphs in Section \ref{sec:IPA}.
			
			Marti et al.~\cite{MARTI-14,MARTI-09} approach the problem using a Greedy Randomized Adaptive Search Procedure (GRASP) with Path Relinking. Jha~\cite{JHA-03} used Genetic Algorithms on the problem of road design and returned intermediate solutions as candidate alternate paths. Later, Kang et al.~\cite{KANG-12,KANG-10} used Genetic Algorithms to generate path alternatives designing new highways that intersect with an existing highway network. Yang et al.~\cite{YANG-14} build on this work using a multi-objective model and modifying the Genetic Algorithm. Genetic Algorithms were also used by Zhang and Armstrong~\cite{ZHANG-05} for the multi-objective corridor problem. We did not study genetic algorithms, or GRASP, in this paper in favour of deterministic algorithms.  Future research should explore comparisons between the approaches herein and nondeterministic methods, as well as explore the possibilities of hybrid approaches.
			
			A different method on multi-objective corridors was used by Dell'Olmo et al.~\cite{DELL-05} to find the pareto-optimal set of non-dominated paths. Despite the different objectives we suggest an algorithm of a similar nature in Section \ref{sec:KSP}.
				
			A common strategy is to generate a large set of candidate paths and then reduce this set based on the chosen dissimilarity criteria. Kuby et al.~\cite{KUBY-97} used a minimax method to select the set of paths which maximized the minimum dispersity between the paths, a problem which is known to be NP-hard~\cite{DUARTE-07}. On HazMat transportation Kang et al.~\cite{KANG-13} generate their initial path set using regular $\kk$-shortest-paths algorithm. This is appropriate for their problem since their objective function is based primarily on risk rather than travel costs. This objective function, unlike one based on construction costs, will often produce $\kk$\textsuperscript{th} best paths that are spatially dissimilar.
			
			The Gateway Shortest-Paths method was introduced by Church and Lombard~\cite{CHURCH-92,LOMBARD-93} as a method of generating a set of spatially dissimilar candidate paths. We use a similar path generation method in Section \ref{sec:BDS}. A very recent method was developed by Scaparra et al.~\cite{SCAPARRA-14} using a multi-gateway shortest-path method. This algorithm shows promise in generating a large number of candidate paths, but requires additional time, requiring a shortest-path algorithm to be run for every vertex in the graph. Given the density of vertices in the road design network this approach would take too long to run and would generate more paths than necessary to select reasonable alternatives. 
			
			We introduce our road design model in Section \ref{sec:The Model} and discuss two height restrictions to decrease running time in Sections \ref{sec:HR} and \ref{sec:EHR}. Section \ref{sec:Dissimilarity Constraint}  contains the dissimilarity constraint required for our corridors. Five dissimilar multipath algorithms are presented and their theoretical performance is discussed in Section \ref{sec:Multipath Algorithms}. Numerical results were collected and are summarized in Section \ref{sec:Sample Tests}. Section \ref{sec:Conclusion} contains some concluding remarks and ideas for further improvements in future work.
				
\section{The Model}\label{sec:The Model}
		\subsection{The \texorpdfstring{$\twoDO$}{twoDO} model}		\label{sec:2DO}

			In order to achieve accurate cost assessments and realistic paths we are using a very dense grid with approximately 10 meters between each vertex. 
			However, with the increased detail of the model a new challenge arises: an alignment with a $90\,^{\circ}$ turn (or more) between two vertices will have to take place in a very short space. This abrupt turn will generally violate road safety curvature constraints, making these alignments infeasible. We begin by introducing a constraint that the sharpest turn any road can take is a $45\,^{\circ}$ angle. While this constraint may prove sufficient for some applications, e.g. logging roads, two consecutive $45\,^{\circ}$ turns would still yield an unsafe highway alignment. However, this method is still sufficient for producing an initial corridor in which further curvature constraints can then be applied during the fine-tuning of the horizontal alignment optimization process.

			\begin{figure}[ht]
				\centering
				\includegraphics{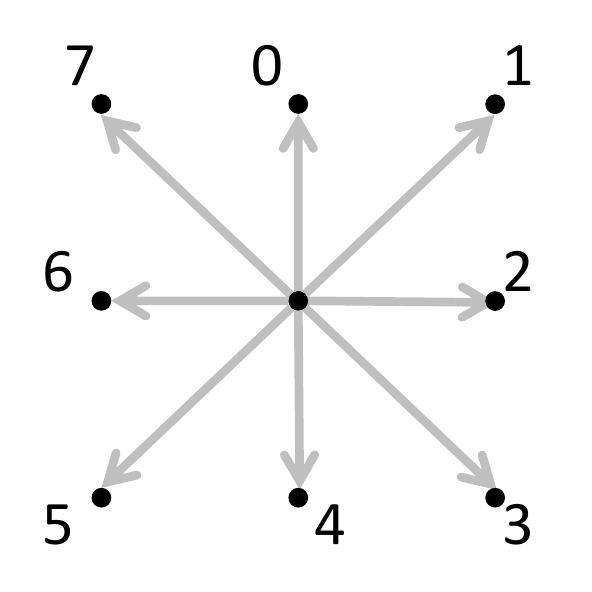} \label{fig:eight directions}
				\caption{The 8 possible $\h$ directions.}\label{fig:directions}
			\end{figure}			
			
			Simply storing the current direction and applying this restriction results in a constrained shortest-path problem. In order to use classic shortest path algorithms, such as Dijkstra's Algorithm, we create a new $\twoDO$' model based on an augmented graph whose shortest-path solution corresponds to a solution to the constrained shortest-path problem. Trietsch~\cite{TRIETSCH-87} used this idea with hexagonal grids to apply curvature constraints to preliminary road designs.

			The augmented graph has eight vertices for each $(x, y)$ point, each one corresponding to an incoming edge orientation, see Figure~\ref{fig:directions}. A simple way of abstracting this information is to consider the direction of travel as another axis, let this be the $\h$-axis, for horizontal  orientation. Each direction can then be assigned a coordinate, 0-7. 	

			Now that we have a new coordinate, we need to modify our original coordinate system. Let $\vertex(\x,\y,\h)$ be a vertex with Euclidean coordinates $(\x,\y)$ and orientation $\h$.			
			 A vertex with orientation $\h$, will then have an outgoing edge to vertices with orientation $\hp$, where
			
			 \[
			 \hp \in
			 \begin{cases}
			 \h - 1,  \\
			 \h,  \\
			 \h + 1,\\
			 \end{cases}
			 \pmod{8},
			 \]

			\begin{thm}
				Finding the shortest-path with the $\twoDO$ model can be solved in log-linear time.
			\end{thm}
			\begin{Proof}
				In the original model, every vertex has at most 8 edges, one to each adjacent vertex (with vertices on the boundary having slightly less edges). So if our map contains $\V$ vertices, we have at most
				\[
					\E \leq 8\V/2 = 4\V \mbox{~edges}.
				\]
				In our new model, we have 8 different possible orientations per vertex. Let $\Vp$ be the new number of vertices, which is now
				\[
					\Vp = 8\V.
				\]
				However, for this model, since we are restricting the possible directions of travel to $45^{\circ}$ angles, we only have 3 possible outgoing and symmetrically 3 incoming edges at each of our augmented vertices, i.e., we have at most
				\[
					\Ep \leq 6\Vp/2 = 24\V = 6\E \mbox{~edges}.
				\]
				Thus, the number of vertices and edges is increased by a constant factor, which means that Dijkstra's Algorithm will still run in log-linear time.
				$\blacksquare$
				\medskip
			\end{Proof}

			An added benefit of this model is that without any further modification, we can now easily improve the accuracy of our cost assessments. Since we are now storing the direction of travel, we can assign different costs to each edge, based not only on their location, but also their orientation. This increases the accuracy of the cost function, as road alignments traveling up a hill vary significantly in cost to those traveling along a hill.

		\subsection{The \texorpdfstring{$\threeDO$}{threeDo} model}
		\label{sec:3DO}

			In order to extend our $\twoDO$ model to three dimensions, we add another axis, call it $\vv$, for vertical orientation. Unlike $\h$, $\vv$ only needs 3 orientations: -1, moving down; 0 maintaining elevation; and 1, moving up. Note $\vertex(\x,\y,\z,\h,\vv)$ the vertex with Euclidean coordinates $(\x,\y,\z)$ and orientation $(\h,\vv)$. A vertex with orientation $(\h,\vv)$ is then a vertex with an incoming edge with orientation $(\h,\vv)$ and an outgoing edge to vertices with orientation $(\hp,\vp)$, where

			\[
				\hp \in
				\begin{cases}
					\h - 1,  \\
	 				\h,  \\
					\h + 1,\\
				\end{cases}
				\pmod{8},
			\]
			\[
				\vp \in
				\begin{cases}
					\min(\vv + 1, 1), \\
					\vv, \\
					\max(\vv - 1, -1), \\
				\end{cases}
			\]

			\begin{thm}
				Finding the shortest-path with the $\threeDO$ model can be solved in log-linear time.
			\end{thm}
			\begin{Proof}
				Similarly to our $\twoDO$ model, the number of vertices and edges increases by a constant factor. The basic 3D model has at most 24 edges per vertex (with boundary vertices having slightly less). If our map contains $\V$ vertices, then we have at most
				\[
					\E \leq 24\V/2 = 12\V \mbox{~edges}.
				\]

				In the new model we have 24 different possible orientations per vertex, which increases our number of vertices to
				\[
					\Vp = 24\V.
				\]
				Since we are restricting the possible directions of travel to $45^{\circ}$ angles, we have up to 9 outgoing and 9 incoming edges at each of our augmented vertices (vertices with $\vv = 0$ have 9 and vertices with $\vv = -1\text{ or }1$ have 6). Which means we have at most
				\[
					\Ep \leq 18\Vp/2 = 9(24\V) \leq 18\E \mbox{~edges}.
				\]
				The number of vertices and edges increases by a constant factor, which means that Dijkstra's Algorithm will still run in log-linear time.
				$\blacksquare$
				\medskip
			\end{Proof}

	In practice Dijkstra's Algorithm is able to find solutions, but requires significant computational time due to the size of the grid used. We discuss three methods of adapting Dijkstra's Algorithm specifically for the road design problem in order to reduce the overall running time required. These modifications can also be used in combination to speed up all of the dissimilar multipath algorithms in Section \ref{sec:Multipath Algorithms} which use Dijskstra's algorithm as their underlying shortest-path algorithm.


			The A* algorithm~\cite{DELLING-09} is one that is commonly used as it maintains the same worst-case complexity as Dijkstra's Algorithm, but in practice it has a faster expected running time. It uses a heuristic for a lower bound of the distance remaining to the destination and modifies the arc weights accordingly. For our application we are using the cost of paving a straight road to the destination. When used in the Bidirectional Selection Algorithm presented in Section \ref{sec:BDS} we modify the heuristic using the formula given by Ikeda et al.~\cite{ikeda94}, which provides the best known running time without losing global optimality.

		\subsection{Cost Formulation}

In this model costs are assigned to each edge in the graph. Depending on the information available, a large variety of costs could be incorporated into the edge cost such as paving, earthwork, land acquisition, user costs, expected accident rates, etc.  Some costs, such as expected accident rates or bridge construction, may be more complicated to quantify as the cost cannot be determined without considering the surrounding edges used in the road.

For the numerical results in Section~\ref{sec:Results}, we used only earthwork and paving costs.  This approach was adopted to make edge costs easy to compute.  The paving costs are simply proportional to the length of the edge, while the earthwork costs are based on the volume of earth excavation or embankment. Earth excavation and embankment was calculated using the height of the edge relative to the elevation profile of the ground. Namely, the ground is a piecewise linear function while the edge is linear so for each linear piece of the ground, we compute the area difference with the edge and multiply by the width of the road to obtain the volume. While changing cost structures may alter the final solutions determined, we believe that algorithmic performance would only be minimally effected.

		\subsection{Simple Height Restriction (HR)}
		\label{sec:HR}

			We note that it is more expensive to build roads that are far above or below the ground than to build roads along the ground. This trend is not exploited by Dijkstra's Algorithm which wastes time searching unrealistic paths. We introduce a simple novel constraint that restricts how far from the ground the algorithm can search. We note the following parameters
			\begin{param}
				\item $\Hm $  maximum displacement allowed from the ground,
				\item $\RR$  radius checked around a given point.
			\end{param}

			In the event the ground changes elevation abruptly, we may need to build a slope to allow our road to change from one elevation to the other. As a result, for these parts of the map we will need to allow the algorithm to search farther away from the ground, which is why we include our radius $\RR$.
				 Let $Z(x,y)$ be the elevation of the ground at $(x,y)$.
			Then for any position $(\bar{x},\bar{y})$, a maximum distance above the ground, $\delta h_+(\bar{x},\bar{y})$, is computed by
			\[
			\delta h_+(\bar{x},\bar{y}) = \max(\max\{Z(x_i,y_j)\}, Z(\bar{x},\bar{y}) + \Hm) - Z(\bar{x},\bar{y}),
			\]
			where
			\[
				Z(x_i,y_i) = \left\{ (x_i,y_j) ~:~
				\begin{array}{l}   \bar{x} - \RR \leq x_i \leq \bar{x} + \RR \\
					  \bar{y} - \RR \leq y_j \leq \bar{y} + \RR\\
				\end{array}\right\}.
			\]
			Symmetrically, we have the maximum distance below the ground
			\[
			\delta h_-(x,y) = \min(\min\{Z(x_i,y_j)\}, Z(\bar{x},\bar{y}) - \Hm) - Z(\bar{x},\bar{y}).
			\]
			
			If $\RR$ is set to 0, then only vertices within $\Hm$ of the ground will be searched, which imposes the height restriction. The value of $\RR$ increases the radius searched at each point when checking to see if the ground elevation abruptly changes. If the ground within the radius extends beyond  $\Hm$, then the height restriction is relaxed to allow these points to be connected.

			As $\Hm$ and $\RR$ decrease, significant reductions in computation time can be observed. However, one must be careful to not over-constrain the problem, which could remove desirable roads from being considered. 
	
		\subsection{Expanding Height Restriction (EH)}
		\label{sec:EHR}

		The Expanding Height Restriction method was designed to initially only consider a small subset of vertices that are near the ground. A set of rules were devised to expand this subset when necessary to ensure that the problem was not over-constrained, removing promising candidate paths.
		
		In particular there are two cases where building a road that is not near the initial ground elevation may have been cheaper. The first is when a cliff steeper than the maximum permissible road grade needs to be climbed, which may require a ``ramp''-like structure to build the road. The second corresponds to when it is cheaper to build a straight road through an obstacle rather than to incur the added pavement costs of building a longer road to avoid the obstacle.
		
		In numerical experiments (see Section~\ref{sec:Sample Tests}), these rules had no effect on solution quality, but significantly slowed solution time.

		\section{Dissimilarity Constraint}
		\label{sec:Dissimilarity Constraint}
		
			We use an area constraint between each of the $\kk$ alternate paths selected by the algorithms to ensure that the alternate paths are not small deviations of the globally optimal path.
		
\begin{figure}[ht]
				\includegraphics[scale=.38]{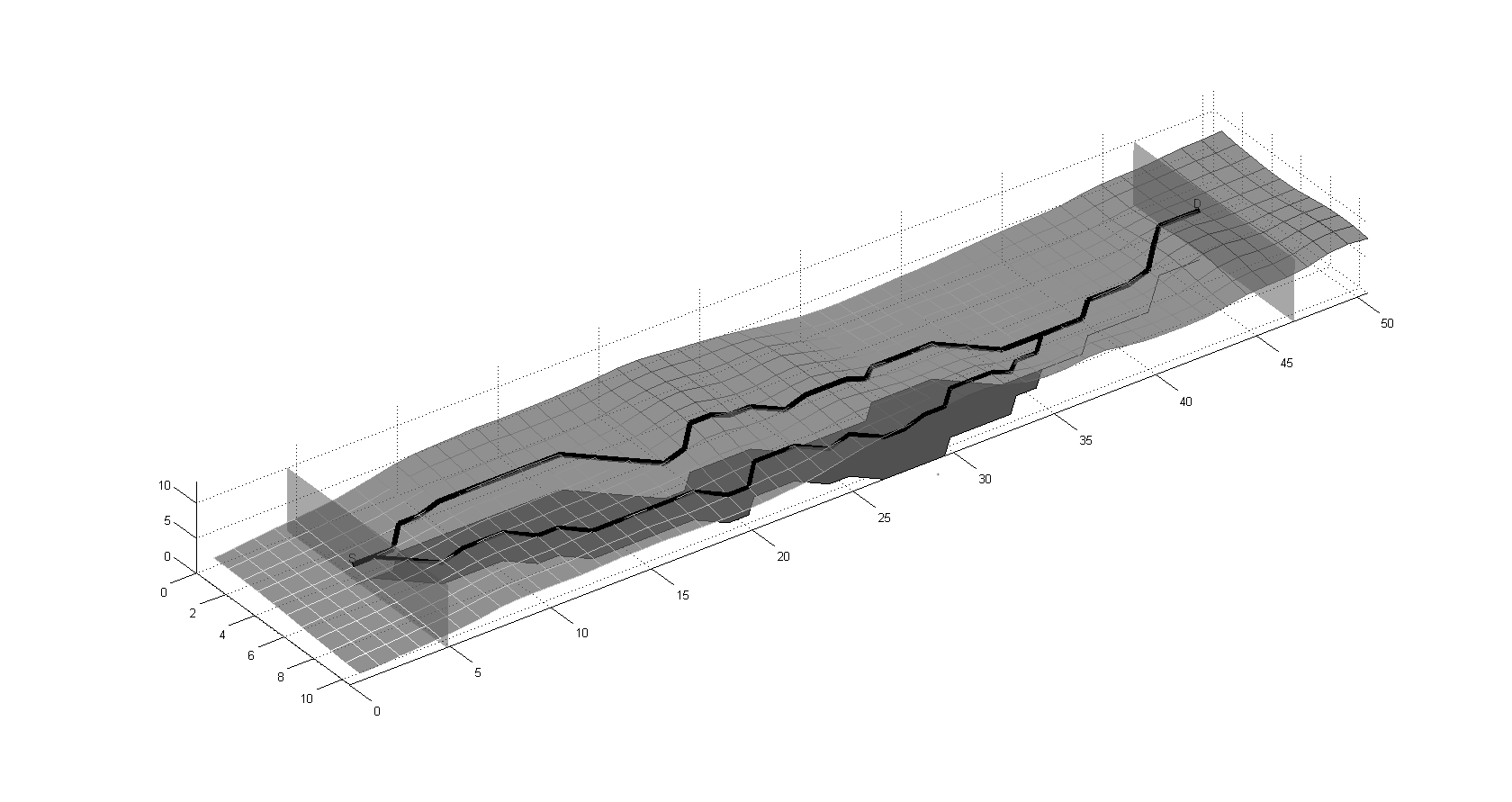}
				\centering
				\caption{The projection of the area between two paths onto an $\x$-$\y$ plane, bounded by the endpoints of the two paths. The percentage of area is then calculated by dividing the area of the projection, by the area of the corridor between the two gray walls shown.}
				\label{fig:areaCalc}
			\end{figure}

			\begin{df}
				We denote by $\areaDiff(\Pathp,\Pathq)$ the percentage area difference of paths $\Pathp$ and $\Pathq$. The percentage area difference between two paths  is found by first projecting the paths onto an $\x$-$\y$ plane, computing the area between the two paths, and then dividing by the area of a rectangle whose width is equal to the width of the map and whose length is equal to the straight-line distance between the endpoints of the paths. See Figure \ref{fig:areaCalc}.
			\end{df}

			We denote by $\minDiff$ the minimum percentage area difference required between each of the alternate paths selected. 	
			
			\begin{lem}
				\label{lem:areaCalc}
				The worst-case time complexity to compute the area between two paths is $\bigO(\PathLength)$, where $\PathLength$ is the number of vertices in the longest path found. Note that $\PathLength \ll n$, where $n$ is the number of vertices.
			\end{lem}
			
			\begin{Proof}
				The area computed between two paths is computed analogously to the integral of the absolute difference of the area under each path. This requires passing the length of the paths three times: once each to compute the area under each vertex and once more to compute the absolute difference. Since each pass requires $\bigO(\PathLength)$ operations, the total time complexity is $\bigO(\PathLength)$.
				$\blacksquare$
				\medskip
			\end{Proof}
			
			This dissimilarity constraint was selected both for its relatively low computation time and its simplicity in implementation. While there are many other possible choices for dissimilarity metrics and this may not be the best one, this was believed to be an adequate metric to enforce spatial path dissimilarity. The constraint is related to the width of the map to help avoid over-constraining the problem. Should the dissimilarity constraint be set too high, most problems will become infeasible. By incorporating the width of the corridor into the constraint, future work can be done to determine a range of values which will typically yield feasible solutions with maximal dissimilarity.
		
	\section{Dissimilar Multipath Algorithms}
	\label{sec:Multipath Algorithms}
			
		This Section presents the dissimilarity criterion used by the five dissimilar multipath algorithms presented in the following subsections. Pseudo code algorithms are given and their theoretical performance is discussed.

		\subsection{Sensitive Elimination Method (SE)}
		\label{sec:SE}
	
	 		The Sensitive Elimination Method is an original algorithm similar in concept to deletion algorithms for finding the $k$-shortest paths~\cite{BRANDER-95,LAWLER-72,YEN-71}. Instead of removing each of the edges in the optimal path, this method seeks to identify a promising edge to cut. The Sensitive Elimination Method begins by computing the optimal path using Dijkstra's Algorithm. Next, some ``sensitivity analysis'' is performed for each edge in the optimal path, to determine which edge is the most sensitive. Then that part of the map is removed from the set of feasible edges and the shortest-path algorithm is again run on this reduced map. We use the following notation
			\begin{param}
				\item $\ww \geq 1$ - sensitivity width,
				\item  $\kk > 1$ - target number of dissimilar paths,
				\item $\minDiff$ - minimum area difference required from every other path  as a percent,
				\item $\maxDiff$ - maximum price difference from the optimal path, as a percent.
			\end{param}

			The sensitivity of a given incoming edge is computed by finding the difference in price of building an edge with the same vertical position and orientation, $(\z, \h, \vv)$, but shifted to the left and right by $d = 1 , 2, ... \ww $. This then produces an array of cost differences to the right and an array of cost differences to the left. Then, the sum of the absolute value of each array is multiplied to give a single numerical representation of the sensitivity of that edge.
		
			The cost model in this paper uses earthwork costs and paving costs. Since the segments to the left and right are the same length, the paving costs will be the same. The difference in costs are based on the earthwork costs, which in turn is based on the displacement of the road segment from the ground. For simplicity, instead of computing the difference in edge costs, we use a simple surrogate and compute the difference in elevation at the vertex to which the edge points.
			
			\begin{figure}[ht]
				\centering
				\subfloat[Part 1][The $x$-$y$ plane.]{\includegraphics[scale=0.7]{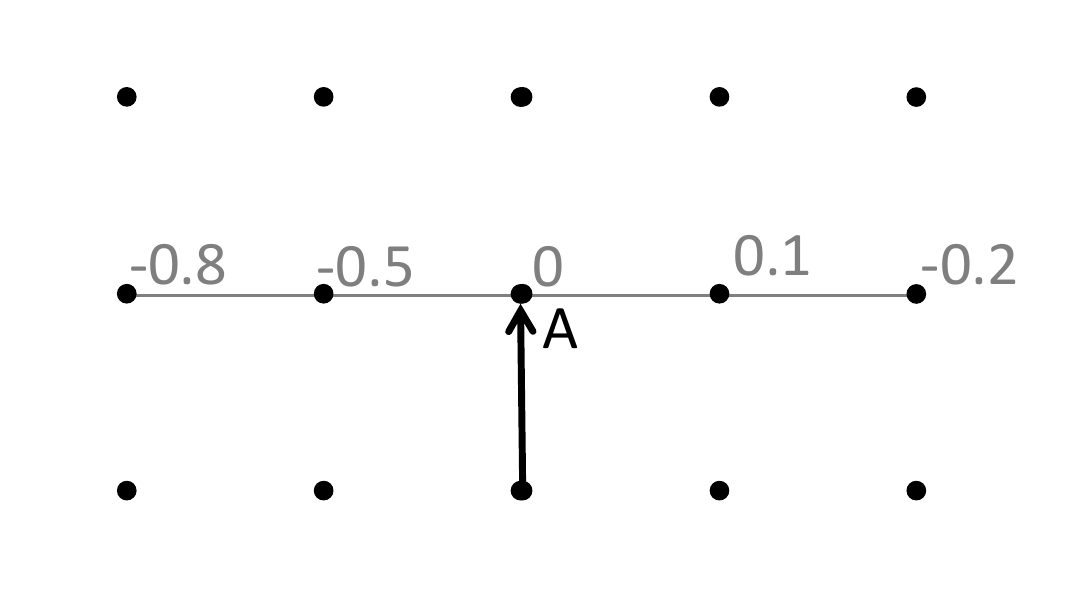} \label{fig:sensitivity-xy}}
				\subfloat[Part 2][The plane formed by the $z$-axis and the grey line in \subref{fig:sensitivity-xy}.]{\includegraphics[scale=0.7]{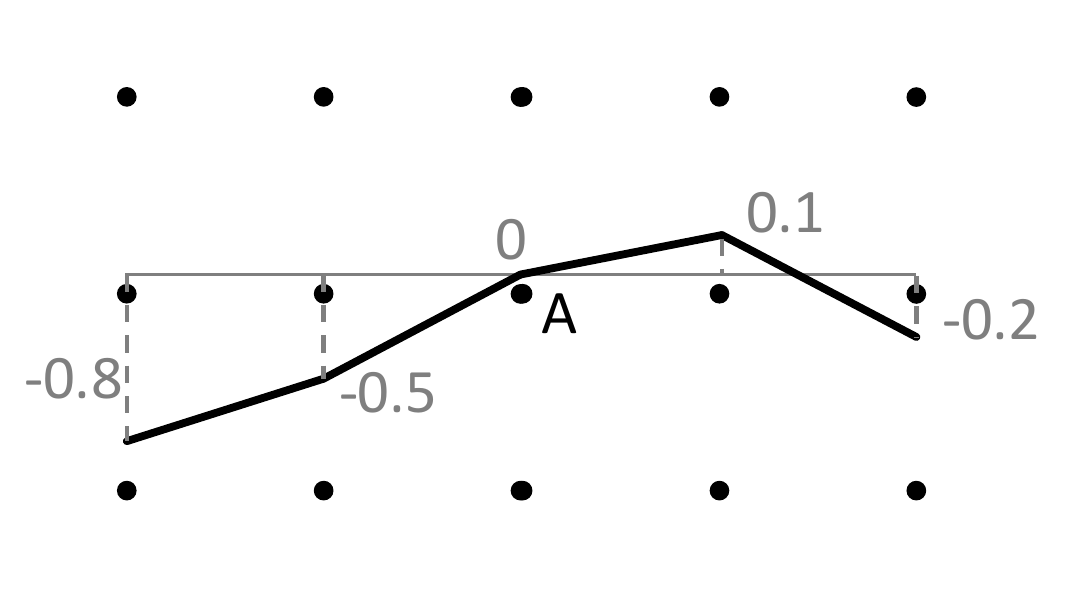} \label{fig:sensitivity}}
				\caption[]{The black line is the elevation profile of the ground to the left and right of vertex $\A$. The solid gray line shows the height of the ground at $\A$ and the doted gray lines show the elevation differences to the left and right of $\A$.}
			\end{figure}

			\begin{ex}
				Consider computing the sensitivity with $\ww = 2$ of the edge going into vertex $\A$. Let the black line in Figure \ref{fig:sensitivity} represent the elevation profile to the left and right of vertex $\A$. The elevation differences to the left and right are $[-0.8, -0.5]$ and $[0.1, -0.2]$ respectively. Taking the absolute value and summing the elements of the arrays gives $1.3$ and $0.3$ respectively. Finally, we multiply these scores to get a raw index of $0.39$, which can be compared with other sensitivity scores. We note that in this example the left side is considered more sensitive than the right, but the final sensitivity index is closer to that of the right. We chose to multiply the sensitivity values of the left and the right so that edges that are sensitive on both sides are selected first.
			\end{ex}

\begin{figure}[ht]
				\includegraphics[scale=0.3]{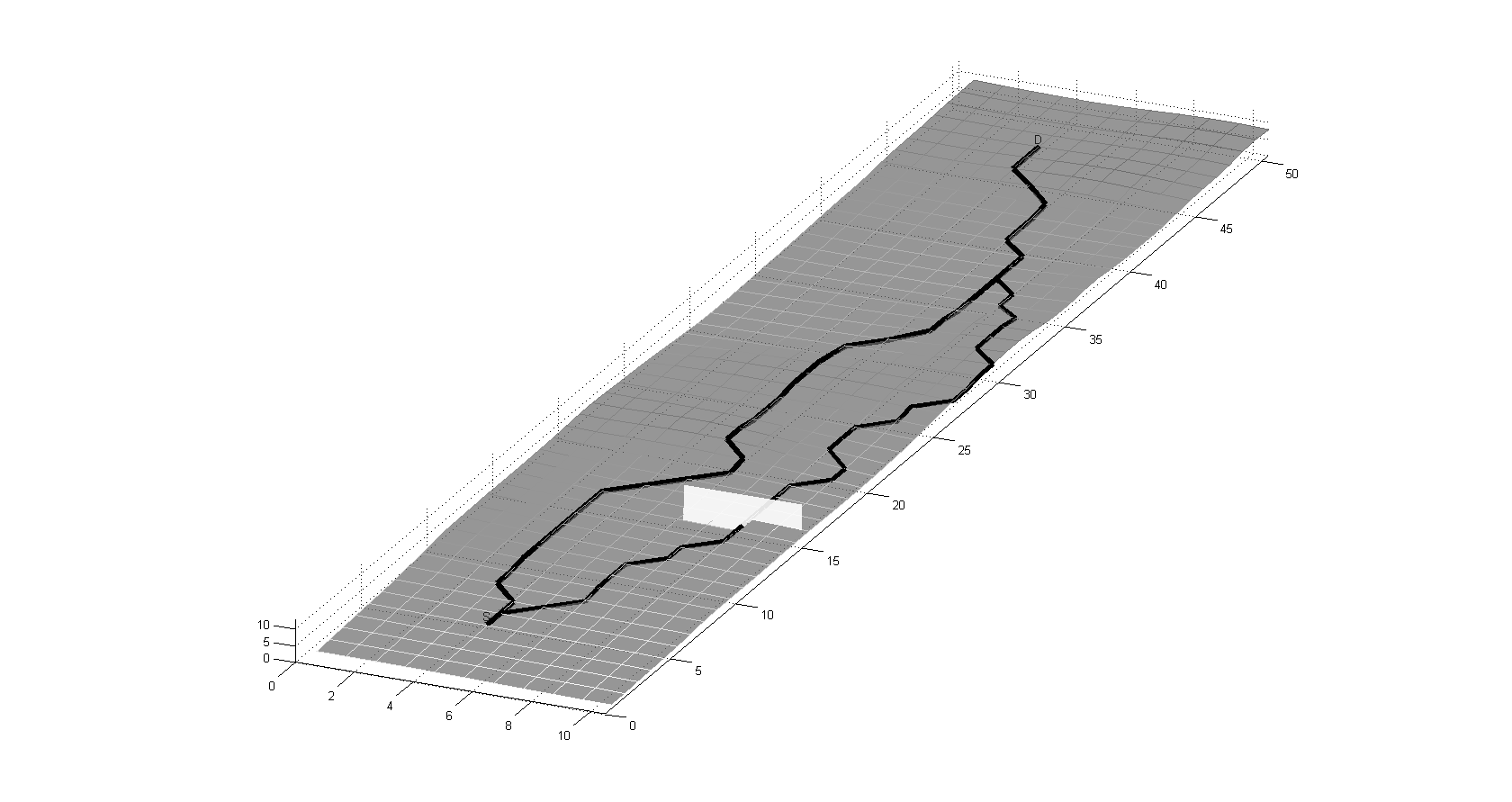}
				\centering
				\caption{Visualization of the Sensitive Elimination Algorithm. The optimal path's most sensitive edge was chosen and the white wall shows the part of the map that was removed, making it impassable. The second path, on the left, is the resulting alternate path found.}
				\label{fig:pathCut}
			\end{figure}

			For a given optimal path, let $\A$ be the vertex pointed to by the edge most sensitive to local perturbations. Instead of only removing $\A$, we also eliminate all of the vertices with $(\x,\y)$ coordinates equal to those of the edges to the left and right of $\A$ that were used to compute the sensitivity of $\A$. By doing this, we have then created a wall in our map centered around the most sensitive vertices of our original path, through which any subsequent alternative paths cannot pass. See Figure~\ref{fig:pathCut}.

			\begin{alg}
				whil\=e we have not yet found $\kk$ paths\+\\
					Compute the optimal path.\\
					if n\=o path was found\+\\
						Replace the edges that were cut from the previous iteration.\\
						Select the next most sensitive, untried edge of the previous path found.\\
						Remove it and those to the left and right.\-\\
					else\=if the new path is too expensive\+\\
						Stop. subsequent paths will also be too expensive.\-\\
					else\= if the new path is too similar to one of the old paths\+\\
						Replace the edges that were cut from the previous iteration.\\
						Select the next most sensitive, untried edge of the previous path found.\\
						Remove it and those to the left and right.\-\\
					else\=\+\\
						Save the newly found path.\\
						Compute the sensitivity of each edge.\\
						Select the most sensitive edge. \\
						Remove it and those to the left and right\-\\
					end\-\\
				end
			\end{alg}

			One of the problems with this method, is that it may require computing many paths that do not get used, a fact that is reflected in the worst-case time complexity.

			\begin{prop}
			\label{prop:SEcomplexity}
				The worst-case time complexity for the Sensitive Elimination Algorithm is  $\bigO(\kk \PathLength n \log(n) + \kk^2 \PathLength^2)$, where $n$ is the number of vertices, $\kk$ is the number of paths to be found and $\PathLength$ is the number of vertices of the longest path considered, with $\PathLength \ll n$.
			\end{prop}

			\begin{Proof}
 				The algorithm first finds the optimal path containing up to $\PathLength$ vertices. It is possible for all but one cut to result in either an infeasible map, or an alternate path that is too similar to another path. This can then result in Dijkstra's Algorithm being run up to $\PathLength$ times to find a single feasible alternative path. Since we need to find $\kk -1 $ alternate paths, we may need to run Dijkstra's Algorithm $\bigO(\PathLength (\kk - 1))$ times, assuming that only the $\PathLength$\textsuperscript{th} iteration results in an acceptable path each time.

				The predicted best cut is selected from each of the $\PathLength$ choices, which requires $\PathLength$ operations to select the maximum value. Dijkstra's Algorithm needs to be run for each attempted cut. Dijkstra's Algorithm requires $\bigO(n \log(n))$, giving us $\bigO(n \log(n) + \PathLength)$ to generate each possible alternate path. Each time a new path is found it needs to be compared with the $\bigO(\kk)$ other paths, requiring $\bigO(\kk \PathLength)$ operations (see Lemma \ref{lem:areaCalc}). Combining these gives us $\bigO(n \log(n) + (\kk + 1) \PathLength)$ operations each time a new path is considered.
				
				Since we may need to repeat Dijkstra's Algorithm at most $\PathLength (\kk - 1)$ times, this yields a total time complexity of $\bigO(\PathLength (\kk - 1)(n \log(n) + (\kk + 1) \PathLength))$, or more simply $\bigO(\kk \PathLength n \log(n) + \kk^2 \PathLength^2)$.
				$\blacksquare$
				\medskip
			\end{Proof}

		\subsection{Iterative Penalty Adaptation (IPA)}
		\label{sec:IPA}
		
			A common method for finding alternative paths is to iteratively apply cumulative penalties along the shortest-paths~\cite{AKGUN-00,JOHNSON-92,ROUPHAIL-96}. First, a shortest-path algorithm is run to find the optimal solution. Weights along the shortest-path are then increased by a penalty and the shortest-path algorithm is run again to find a second alternative path. This process is repeated iteratively to find the desired number of paths.
	
			Due to the dense nature of our graphs if we were to apply this method directly we would find that the secondary path found would often simply be the original path shifted ten meters to the left or right to avoid the penalty, which does not provide us with sufficiently distinct alternate paths. We adapt existing methods by not only increasing the weight of the optimal path, but by also increasing the weight of a corridor-like buffer around the optimal path to discourage the alternative paths from being chosen too close to the original paths. This is achieved by adding a weight to the optimal path and decreasing that weight linearly on both sides of the path till it reaches zero. The effect of the buffer is to consider that any minor variation of the optimal path can be built by the engineer on the ground since our goal is to identify the corridor, not compute a precise path. Hence, the buffer forces any new solution to be clearly distinct.

 			Iterative Penalty methods inherently have many variable options that will determine the performance of the algorithm. One could use either an additive or a multiplicative penalty. For simplicity we have opted to penalize all of the edges near the original paths with an additive penalty. We used a decaying penalty - that is, one with less weight farther away from the original paths. For simplicity we used a triangular decay shape with a fixed ratio of length to height and introduce the following notations
			\begin{param}
				\item $\penaltyWidth$ - the initial penalty width,
				\item $\penaltyMax$ - the algorithm terminates when $\penaltyWidth$ exceeds this value,
				\item  $\kk > 1$ - number of paths,
				\item $\minDiff$ - minimum area difference required from every other path  as a percent,
				\item $\maxDiff$ - maximum price difference from the optimal path, as a percent.
			\end{param}
		
			Each iteration of the algorithm involves running a shortest-path algorithm and applying the corridor weights. The new path found will not always satisfy the price and area constraints. If this is the case we will either decrease or increase the penalty width, $\penaltyWidth$. We maintain a bracket beginning with a lower bound of zero, initially without an upper bound. When $\penaltyWidth$ is increased we will initially double its value until it needs to be decreased, providing an upper bracket. Each time $\penaltyWidth$ is changed the distance between it and the appropriate bracket is halved and the new bracket values are updated accordingly.

			\begin{alg}
				whil\=e we have not yet found $\kk$ paths\+\\
					compute the optimal path.\\
					if t\=he new path is too expensive\+\\
						try the middle of the $\penaltyWidth$  bracket\\
						if w\=e have already tried the new $\penaltyWidth$\+\\
							stop\-\\
						end\-\\
					else\=if the new path is too similar\+\\
						if w\=e have an upper bound for the $\penaltyWidth$ bracket\+\\
							try the middle of the $\penaltyWidth$ bracket\-\\
						else\=\+\\
							double the $\penaltyWidth$\-\\
						end\\
						if w\=e have already tried the new $\penaltyWidth$, or $\penaltyWidth > \penaltyMax$\+\\
							stop\-\\
						end\-\\
					else\=\+\\
						add the new path to the set of $\kk$ paths\\
						clear data about which values of $\penaltyWidth$ have been used\-\\
					end\\
					save that the current value of $\penaltyWidth$ has been used\-\\
				end\\
			\end{alg}
			
			\begin{prop}
			\label{note:IPAcomplexity}
				The Iterative Penalty Adaptation method has a worst-case running time of $\bigO(\kk^2 n \log(n) \log(\penaltyMax) + \kk^3 \PathLength \log(\penaltyMax))$.
			\end{prop}
			
			$\Proof$
			Each shortest-path algorithm iterations takes $\bigO(n \log(n))$ operations. Each time a path is found it must be compared to the $\bigO(\kk))$ other paths, requiring $\bigO(\kk \PathLength)$ operations (see Lemma \ref{lem:areaCalc}). This gives a time complexity of  $\bigO(\kk n \log(n) + \kk^2 \PathLength)$ for each iteration.
				
			However, since each iteration may not result in a feasible path the algorithm may have to be run again. Initially the penalty width, $\penaltyWidth$, does not have an upper bracket. As long as the path found is always too similar to the original paths it will continue to double $\penaltyWidth$. Since $\penaltyWidth$ is doubled it will run at most $\bigO(\log(\penaltyMax))$ times before reaching $\penaltyMax$ and terminating, or decreasing and setting an upper bound. The upper bound will be bounded by $\penaltyMax$ and since each time the size of the bracket is halved, the algorithm will find the next path, or run out of new $\penaltyWidth$ values, in at most $\bigO(\log(\penaltyMax))$ iterations.
				
			Each time a new path is found the bracket values are cleared and the data about which $\penaltyWidth$ values have been tried before are reset. This means that the algorithm can spend up to $\bigO(\log(\penaltyMax))$ iterations again to find each of the $\kk$ paths, requiring a total of $\bigO(\log(\penaltyMax))$ iterations.
		
			Since each iteration requires $\bigO(\kk n \log(n) + \kk^2 \PathLength)$, this yields a total time complexity of $\bigO(\kk^2 n \log(n) \log(\penaltyMax) + \kk^3 \PathLength \log(\penaltyMax))$.
			$\blacksquare$
			\medskip 			
 			
		\subsection{\kkk-shortest-paths Adaptation (KSPA)}
		\label{sec:KSP}
			The $k$-shortest-paths Adaptation~\cite{ksp} (KSPA) algorithm was designed  to take advantage of work that has already been done by computing all $k$ paths simultaneously, storing common information for each path only once. It is based on an adapted form of a $k$-shortest-paths algorithm that can be derived from a generalization of Dijkstra's Algorithm.

			When a suboptimal path is found to a vertex, instead of discarding it, we store both the information of the optimal path and the suboptimal path, until we have at most $\kappa$ paths to each vertex, one optimal, and $\kappa-1$ suboptimal paths. In order to maintain the desired properties of our alternate paths, we enforce a restriction that the area between each suboptimal path to the same vertex is at least $\minDiff$. 
			
			Once we have found our optimal path, we can then set the termination condition to be when we have found $\kappa = \kk$ paths to the destination, or when the cost of new paths is more than $\maxDiff$ the price of the optimal path. We note the following parameters
			\begin{param}
				\item $\kappa = \kk$ - number of paths found to each vertex,
				\item $\maxDiff$ - maximum price difference from the optimal path as a percent,
				\item $\minDiff$ - minimum area difference required between every path as a percent.
			\end{param}

			Let
			\begin{tightenumerate}
				\item $\activeM$ be a heap containing vertices that are adjacent to the known shortest-path tree from the source,
				\item $vertex_s$ be the source vertex,
				\item $vertex_d$ be the destination vertex,
				\item $paths_{s-d}$ be a collection of found paths from the source to the destination,
				\item $dim_y$ be the maximum $\y$ value (width of the corridor).
			\end{tightenumerate}
			\medskip

			Also let each vertex have $\kappa$ paths to it, each associated with its own cost. (Rather than each vertex storing only the single best path and cost.)

			\begin{alg}
				push $\sVertex$ onto the $\activeM$ with a cost of 0\\
				whil\=e we have not yet reached the destination:\+\\
					$\current$ = cheapest vertex in $\activeM$\\
					for \=each $\neighbour$ of $\current$\+\\
						if t\=he new path costs within $\maxDiff$ of the cheapest path to $\neighbour$ \+\\
							if t\=he new path is more than $\minDiff$ from other paths to $\neighbour$\+\\
								if w\=e have not yet found $\kappa$ paths to $\neighbour$\+\\
									push $\neighbour$ onto $\activeM$ with its updated cost\-\\
								else\=if this path is cheaper than the most expensive path to $\neighbour$\+\\
									replace the most expensive path and push this one onto $\activeM$\-\\
								end\-\\
							else\=if this path is similar to exactly one of the other paths\+\\
								if t\=he new path is the cheaper of the two\+\\
									replace the old path and push the new one onto $\activeM$\-\\
								end\-\\
							end\-\\
						end\-\\
					end\-\\
				end\\
			\end{alg}

			\begin{figure}[ht]
				\includegraphics[scale=0.8]{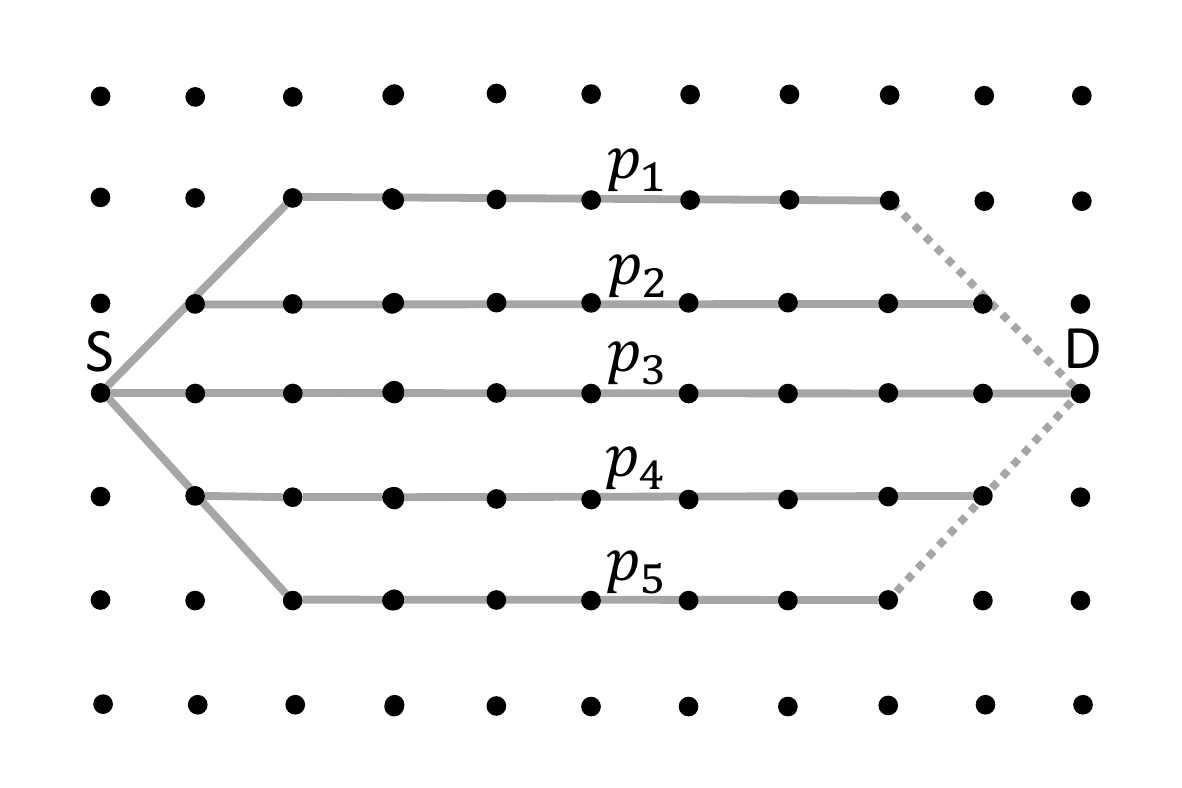}
				\centering
				\caption{A uniform cost grid illustrating the behaviour of a $k$-shortest-path algorithm.}
				\label{fig:KSP-Similar}
			\end{figure}
			
			One limitation of the $k$-shortest-paths Adaptation algorithm is the behaviour that is best illustrated on near-uniform cost grids as may be found in some prairies. As an academic example, consider the five paths $\Path1$, ... $\Path5$ on a uniform cost grid, as seen in Figure~\ref{fig:KSP-Similar}. Let $\B$ be the parent of $\A$ on this path. The path $\Path3$ is the cheapest path. Paths $\Path2$ and $\Path4$ are too similar to $\Path3$ and so will be rejected. Paths $\Path1$ and $\Path5$ are dissimilar from $\Path3$ and so would be acceptable alternate paths, however they are too similar to $\Path2$ and $\Path4$ respectively and so will also be rejected. When each path is very similar to the paths on either side of it, the algorithm is only ever to find one path to each vertex and hence will only find one possible road alignment.

			\begin{prop}
			\label{prop:KSPAcomplexity}
				The $k$-Shortest-Paths Adaptation Algorithm achieves worst-case bounds of $\bigO(\kappa^2 \PathLength n \log(n) )$, where $n$ is the number of vertices, $\kappa$ is the number of paths found to each vertex, and $\PathLength$ is the number of vertices of the longest path considered, with $\PathLength \ll n$.
			\end{prop}

			\begin{Proof}
				Since we are now finding at most $\kappa$ paths to every vertex, instead of each vertex being used in at most 1 path, we can now have each vertex used in at most $\kappa$ paths. This is then similar to having $\kappa n$ vertices, which means that we will have $\bigO(\kappa n \log(\kappa n))$ iterations of the algorithm.
				
				In each iteration of the algorithm we need to compare the new path we have found with up to $\kappa$ other paths. From Lemma \ref{lem:areaCalc} we see that each iteration will then have an additional $\bigO(\kappa \PathLength)$ operations. This gives us a total worst-case time complexity of $\bigO(\kappa^2 \PathLength n \log(\kappa n) )$, or more simply $\bigO(\kappa^2 \PathLength n \log(n) )$.
				$\blacksquare$
				\medskip
			\end{Proof}

		\subsection{Bidirectional Selection Method (BDS)}
		\label{sec:BDS}

			The Bidirectional Selection Method (BDS) is a simple extension of Bidirectional Dijkstra's Algorithm, similar to the approach used by Lombard and Church~\cite{LOMBARD-93}. Instead of using the regular termination condition for the Bidirectional Dijkstra's Algorithm, we simply continue to grow both ends until we have found $\kk$ dissimilar paths. Alternatively, we terminate the algorithm if the cost of the new paths found exceeds the maximum cost restriction, since we will not find any cheaper paths beyond those.

			The current version of our algorithm  uses a method of path selection that is relatively simple. Each time a new path is found it is compared to the set of accepted paths and is added to the set, replaces one of the paths in the set, or is rejected. A computational shortcut is used by not considering a path again after it has been rejected. While this shortcut may result in a desirable solution being missed, the additional computation time required to process all possible road alignment groupings would be infeasible.
			
			A similar method was proposed to find dissimilar paths in road networks~\cite{ABRAHAM-13,LOMBARD-93}. As they discussed, the $\SD$ paths produced by this algorithm have the specific property that given an arbitrary vertex $\A$ an $\SD$ path is formed by concatenating the cheapest path from $\sVertex$ to $\A$ and the cheapest path from $\A$ to $\dVertex$. The parameters are
			\begin{param}
				\item $\kk > 1$ - number of paths,
				\item $\maxDiff$ - maximum price difference from the optimal path as a percent,
				\item $\minDiff$ - minimum area difference required from every other path as a percent.
			\end{param}

			Let
			\begin{tightenumerate}
				\item $\sVertex$ be the source vertex,
				\item $\dVertex$ be the destination vertex.
			\end{tightenumerate}

			\begin{alg}
				push $\sVertex$ and $\dVertex$ onto their respective heaps with a cost of 0\\
				whil\=e there are still entries in the source or destination heaps\+\\
					grow from whichever side has a smaller heap\\
					if t\=he newly added vertex forms a new path\+\\
						if t\=his new path is more than $\minDiff$ from the other paths selected\+\\
							if w\=e have not yet selected $k$ paths\+\\
								save the newly found path in our set of $\kk$ paths\-\\
							else\=if this path is cheaper than the most expensive path selected\+\\
								replace the most expensive path\-\\
							end\-\\
						else\=if this path is similar to exactly one of the other paths\+\\
							if t\=he new path is the cheapest of the two\+\\
								replace the old path with the newly found path.\-\\
							end\-\\
						end\-\\
					end\\
					if w\=e have not yet found $\kk$ paths and new paths are less than $\maxDiff$ of the cheapest path\+\\
						push the neighbours of the newly added vertex onto the heap, as needed\-\\
					end\-\\
				end\\
			\end{alg}

			\begin{lem}
			\label{lem:bidirectionalBounds}
				The Bidirectional Selection method computes at most $n$ possible alternate paths with worst-case time complexity $\bigO(n \log(n))$.
			\end{lem}
			\begin{Proof}
				For a given vertex, we are computing the cheapest path to the source and the cheapest path to the destination. That is, we compute the cheapest path from the source to the destination, that passes through each vertex. This is done by running Dijkstra's Algorithm twice, once from the source and once from the destination, each taking $\bigO(n \log(n))$. By concatenating these results, we then generate one path for each vertex, taking an additional $\bigO(n)$. This gives a combined worst-case time complexity of $\bigO(n \log(n) + n)$, or $\bigO(n \log(n))$.
				$\blacksquare$
				\medskip
			\end{Proof}

			\begin{prop}
			\label{prop:BDScomplexity}
				The Bidirectional Selection Algorithm has worst-case bounds $\bigO(n \log(n) + \kk \PathLength n)$, where $n$ is the number of vertices, $\kk$ is the number of paths, and $\PathLength$ is the number of vertices of the longest path considered, with $\PathLength \ll n$.
			\end{prop}

			\begin{Proof}
		 		From Lemma \ref{lem:bidirectionalBounds} our algorithm finds at most one different path per vertex, we are then selecting $\kk$ paths from a set of at most $n$ paths. Each new path is compared to our set of up to $\kk$ accepted paths, which, from Lemma \ref{lem:areaCalc}, requires $\bigO(\kk \PathLength)$ operations for each new paths. With $n$ new paths this gives us a total worst-case time complexity of $\bigO(n \log(n) + \kk \PathLength n)$.
		 		$\blacksquare$
				\medskip
			\end{Proof}

		\subsection{BDS-KSPA Hybrid Method}
		\label{sec:KSP-BDS}

			This algorithm combines the two strategies by running the KSPA algorithm from both directions, as is done in the BDS method. We do make note that the number of paths found to a particular vertex from one direction, as used in the $k$-shortest-paths Adaptation algorithm, need not be the same as the total number of paths found.  The value of $\ka$ can be thought of as the number of shortest-path trees made from both the source and the destination, while $\kb$ controls the number of paths that are selected and returned from the paths generated by the shortest-path trees. A value of $\ka = 1$ reduces the algorithm to the Bidirectional Selection method, but a large value of $\ka$ will increase the running time of the algorithm.

			We omit the pseudo code algorithm for this section. The modifications provided in Section \ref{sec:KSP} do not affect the new termination and path selection process described in the pseudo code of \ref{sec:KSP-BDS}.

			\begin{prop}
				The BDS-KSPA hybrid method has $\bigO(\ka^2 \PathLength n \log(\ka n) + n \ka \kb \PathLength)$ as a worst case time complexity, where $\PathLength$ is the number of vertices in the longest path considered, with $\PathLength \ll n$.
			\end{prop}
			
			\begin{Proof}
				Similarly to the KSPA algorithm each shortest-path tree generation requires $\bigO(\ka^2 \PathLength n \log(n))$ operations as seen in Proposition \ref{prop:KSPAcomplexity}.
				
				Since we are now finding at most $\ka$ paths to each vertex we are generating a set of $\bigO(\ka n)$ paths. Since this algorithm uses the same path selection process as the BDS method each path will require $\bigO(\kb \PathLength)$ operations (see Proposition \ref{prop:BDScomplexity}). This gives a total of $\bigO(n \ka \kb \PathLength)$ operations for the path selection process.
				
				Combining these results gives us a worst-case time complexity of $\bigO(\ka^2 \PathLength n \log(n) + n \ka \kb \PathLength)$ for the BDS-KSPA hybrid method.~$\blacksquare$
				\medskip
			\end{Proof}

\section{Numerical Tests}
\label{sec:Sample Tests}

	Numerical tests were run to compare algorithm quality in terms of time required and ability to find a valid solution. We look at which algorithm is most consistently able to find spatially dissimilar paths that are near the cost of the globally optimal path. The running time is also compared to select the fastest algorithm with the best results. The numerical tests are also used to measure the improvement in the running time gained with the height restrictions.
	
	A parameter required for each test is $\minDiff$, the difference threshold to state two given paths are `distinctly different'. This value should be selected based upon the number of paths desired to be found, $\kk$. We have chosen $\kk = 3$ and we are using $\minDiff = 12\%$. Another parameter required for both is $\maxDiff$, which is the maximum cost difference allowed in secondary paths, as a percent of the optimal path. We have selected $\maxDiff = 10\%$.
				
	In our numerical results  an algorithm is considered to have found a solution if it meets three criteria. First, we require three paths with one of them being the optimal path; second, the maximum path cost must be within the specified range of the cheapest path; and third, the minimum percentage area requirement must be satisfied.
	
	\subsection{Test Environment}
	\label{sec:Test Sets}

		The algorithms perform differently based on the dimensions of the map (in terms of number of vertices). For both of our test sets we are using approximately 10 meters between each horizontally-adjacent vertex and 1 meter between each vertically-adjacent vertex. The first test set has three different map lengths of 40, 80, and 160 vertices, while the second test set has map lengths of 320 and 640 vertices. For each road length we have 7 maps with a length to width ratio of 2:1 and 3 maps with ratios 8:1, giving us a total of 30 maps in test set 1 and 20 maps in test set 2. The initial terrain data was obtained from the USGS National Map Viewer~\cite{nationalMap}. The individual maps were found by sampling mountainous and prairie regions and selecting a variety of different vertical dimensions to ensure a good spread of map types. We provide a detailed breakdown of the terrain features present in each map in Appendix \ref{app:Test Sets}. The exact data used can be obtained by contacting the authors.

			All of the tests for the first test set were run on the Orcinus cluster, a 9600 core available through Westgrid~\cite{westgrid}. Each test was run on a single core of an Intel Xeon X5650 six-core processor, running at 2.66GHz. The different map sizes were allotted different amounts of memory, up to the 24GB of RAM available per node. The second test set was run on Grex, a 3792 core cluster available through Westgrid. Again, each test was run on a single core of an Intel Xeon X5650 six-core processor, running at 2.66GHz. Each node had up to 96GB of RAM available. All of the code was written in MATLAB R2013b. Note that the algorithms are serial; the only reason we performed the tests on a cluster was to speedup running the algorithms on the full test set.

			We have selected five algorithms for testing in this paper, and three modifications that can be made to each to improve the performance. We include results for each of the basic algorithms without modification. We then selected the two best algorithms, and present the improvements gained by the three modifications. First they are combined with the heuristic described in Section 
\ref{sec:3DO}, adapted from the A* algorithm. Next they are combined with both the adapted form of the A* algorithm, and the Simple Height Restriction from Section \ref{sec:HR}. Further tests were done with both the heuristic from the A* algorithm, as well as the Expanding Height Restriction from Section \ref{sec:EHR}. 
			
			When the algorithms are using the Simple Height Restriction we have set the following parameters: $\RR = 3$ and $\Hm = 1$. The Expanding Height algorithms will be using a value of $\Hi = 0.5$. We used these parameter values as preliminary experimentation indicated they would provide good results.

			The Sensitive Elimination method's performance is related to the choice of $\ww$, which determines how much of the map is removed at each iteration and $\minDiff$. If $\ww$ is too small, then it is unlikely that the alternative paths found will be at least $\minDiff$ different. As a result, we have chosen to use a value that can be calculated by $\ww = \text{round}(\minDiff \cdot w_m - 0.5)$, where $w_m$ is the width of the map.
			
			Preliminary testing done for the Iterative Penalty Adaptation method achieved optimal performance with an initial penalty width value of $\penaltyWidth = 10\%$, which is the value we used for our tests.
			
			For the BDS-KSP hybrid method we used a value of $\ka = 2$, so that at most two paths would be found to each vertex and a value of $\kb = 3$ so that a total of 3 paths would be found.

		\subsection{Results}
		\label{sec:Results}

\begin{figure}[ht]
			\includegraphics[scale=0.8, trim=110 220 110 220,clip=true]{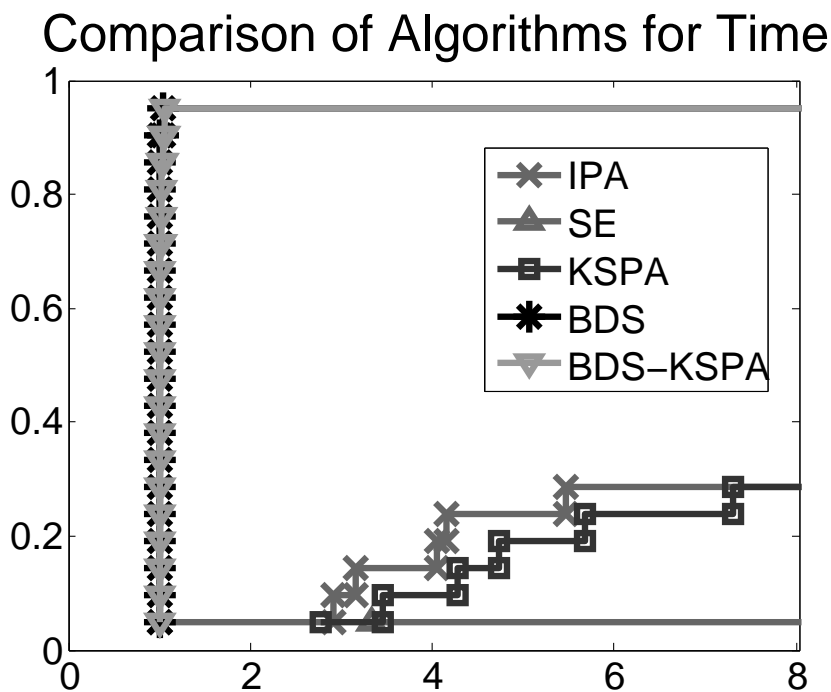}
			\centering
			\caption{The performance profile of all the basic algorithms measured over time.}
			\label{fig:Perf All Algorithms}
		\end{figure}

\begin{figure}[ht]
			\includegraphics[scale=0.8, trim=40 200 40 170,clip=true]{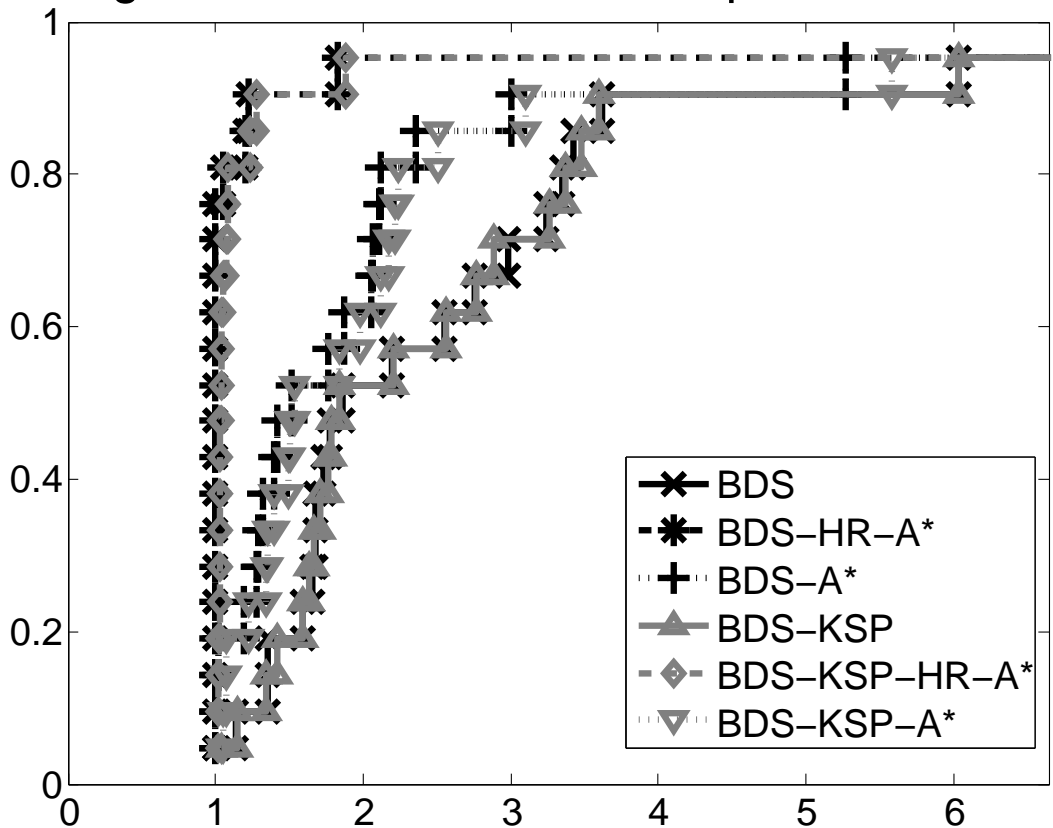}
			\centering
			\caption{The performance profile of the Bidirectional Selection algorithm and the Hybrid Method combined with the Simple Height Restriction and A* component, compared to just the A* component, and the base algorithms.}
			\label{fig:Perf Height Restrictions}
		\end{figure}

		We summarize the results of our numerical tests with the two performance profiles in Figures~\ref{fig:Perf All Algorithms} and~\ref{fig:Perf Height Restrictions}. A performance profile~\cite{DOLAN-02} shows on the $y$-axis the portion of problems that an algorithm was able to solve, while the $x$-axis is the time it took to solve those problems as a ratio of the time it took the fastest algorithm to solve each problem.
		
		We compare our algorithms using the total running time of each algorithm as the basic metric. We can see from our results in Figure~\ref{fig:Perf All Algorithms} that the Bidirectional Selection method was able to solve the most maps in the least time. The next best algorithm was the Iterative Penalty Adaptation algorithm, followed by the $\kk$-shortest-paths Adaptation algorithm.  The Sensitive Elimination method was often unable to find adequate solutions before reaching a predetermined time-out time. We discourage use of the latter three methods, as none of them have shown promising results within reasonable time frames.
		
		In the first test set, the Bidirectional Selection and the hybrid methods were able to find solutions for 20 of the 30 problems. Of the remaining 10 problems, 9 were unsolved by any method, and 1 was solved by the Iterative Penalty Adaption method. Since the problems were designed using real-world terrain data and none of the algorithms found a solution, it is possible that the 9 unsolved problems do not have three feasible roads that satisfy the costing and percentage area difference constraints. As such, we have removed the 9 unsolved problems from our data analysis (including performance profiles in Figures~\ref{fig:Perf All Algorithms} and~\ref{fig:Perf Height Restrictions}).

		While all of the algorithms were run again with the heuristic modifications for efficiency, the first three continued to perform poorly so we do not include their results here. We found that in each case the height restrictions show improvements to the efficiency of the algorithms. For simplicity we omit the results of the Expanding Height Restriction in Figure~\ref{fig:Perf Height Restrictions}, as they provided the same quality of solutions as the Simple Height Restriction, but the added overhead meant that the running time was not competitive.
		
		Our fastest algorithm, the Bidirectional Selection method with the Simple Height Restriction and the A* component, reported a solution to 20 of the 30 problems in the first test set. It was able to solve each of these 20 problems in just over five hours, with the smallest problem being solved in only twenty-three seconds. The same method required less than 4 GB with an average of 20 minutes for the problems which had a map length of 40 vertices; those with map length of 80 vertices took 7 GB and an average of 1.5 hours; and the rest with under 19 GB and an average of 2.3 hours, on maps with a length of 160 vertices.
		
		The Bidirectional Selection method with the Simple Height Restriction and A* component was run again for the second test set. It was able to solve nine of the first ten problems with map length of 320 vertices, but only one of the last ten problems with 640 vertices. Maps with length of 320 vertices had an average running time of 15 hours. The largest problem with 320 by 160 by 195 vertices required the most memory at 82.2 GB. The longest running time of all of the problems was a little more than 28 hours. Of the problems with length 640 vertices, only the smallest  one with 640 by 80 by 59 vertices was solved, using 21.2 GB and just over 16 hours. The other nine problems with lengths of 640 vertices hit the memory limit of 92 GB.
		
		Note that these timings should not be viewed as absolute since coding the algorithms in C instead of MATLAB will greatly reduce the running times. These results show the relative performance of each algorithm, in order to select the best one for further refinement.
		
		\subsection{Example Solution}
		\label{sec:Example Solution}
		
\begin{figure}[H]
				\includegraphics[scale=0.43, trim=80 150 100 110,clip=true]{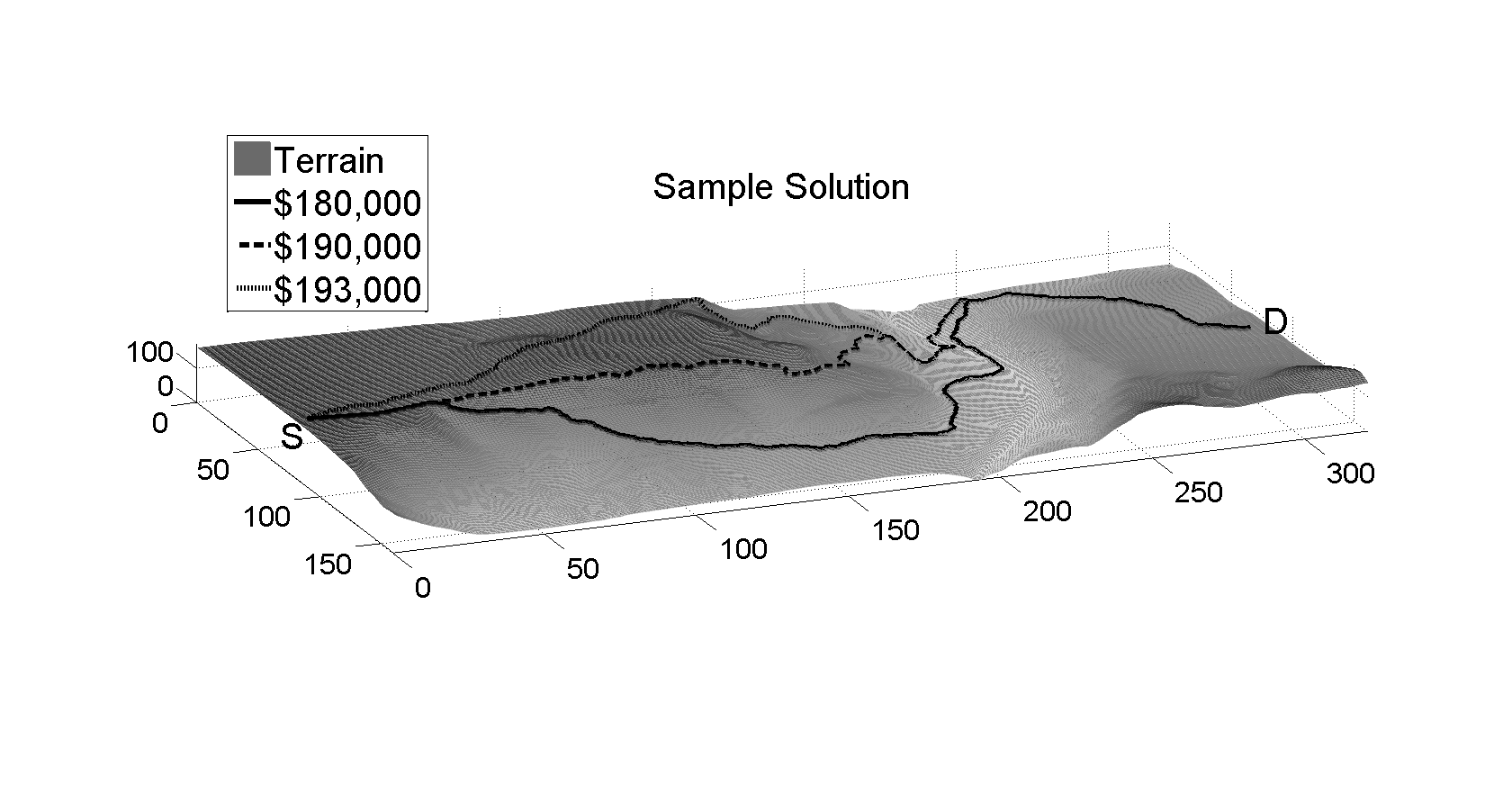}
				\centering
				\caption{The three paths returned by the Bidirectional Selection Method on a map with 320 by 160 vertices (3.2~km by 1.6~km). This terrain corresponds to Test 9 in the second test set.}
				\label{fig:Sample Solution}
			\end{figure}

		In this section we present a sample solution found by the Bidirectional Selection Method on a map with 320 by 160 vertices (3.2~km by 1.6~km), see Figure~\ref{fig:Sample Solution}. The globally optimal path, $\Path1$, was determined to cost approximately 180,000 monetary units. The two alternative paths, $\Path{2}$ and $\Path{3}$ cost 190,000 and 193,000 monetary units, respectively. These costs satisfy the maximum cost constraint as they are both within 110\% of the cost of the globally optimal path, at 105\% and 107\% respectively. The projected area difference between $\Path{1}$ and $\Path{2}$ is 15\%, the area between $\Path{2}$ and $\Path{3}$ is 13\%, and the area between $\Path{1}$ and $\Path{3}$ is 28\%. This meets the minimum area difference constraint which requires each path to be separated by at least 12\% of the area of the map.
		
	\section{Conclusion}
	\label{sec:Conclusion}

		The primary contributions of this paper are the adaptations of several path-finding algorithms to develop dissimilar multipath algorithms. The Sensitive Elimination algorithm in Section~\ref{sec:SE} is an original method belonging to the deletion family of $\kk$-shortest path algorithms. The Iterative Penalty Adaptation algorithm in Section~\ref{sec:IPA} modifies a common technique by adding a corridor to which a penalty is applied. All of the algorithms were modified to include an acceptance test based on the dissimilarity criterion in Section~\ref{sec:Dissimilarity Constraint}. We introduce a possible hybridization of the two methods BDS and KSPA in Section~\ref{sec:KSP-BDS}.

		Our numerical results indicate that, of the algorithms tested, the only reasonable algorithm for practical use is the Bidirectional Selection method using A* and the Simple Height Restriction. The Bidirectional Selection method and its hybrid with the $k$-shortest-paths Adaptation algorithm, were observed to outperform the other methods in both time and quality of solutions. The main restriction of the method is that it only considers paths that are a concatenation of the cheapest path from the source to a node and then the cheapest path from that node, to the destination. A result of this, is that its paths cannot begin by overlapping, then split, merge and split again. This property does not, however, stop the algorithm from producing two candidate paths which cross at an angle. We did not find this property to be a significant problem for our application, since it was still able to generate a large number of candidate paths. In practice, if an algorithm with this path property was unsatisfactory, then a hybrid of the Bidirectional Selection Method and the $\kk$-shortest-paths Adaption algorithm could be used as described in Section \ref{sec:KSP-BDS}. While this does theoretically increase the range of paths that can be found our numerical results showed little difference in the quality of the solutions between the two algorithms.
		
		Current implementations of the Bidirectional Selection method do not consider a path once it has been rejected. Future work may be able to improve the quality of the solutions by relaxing this condition. Such modifications may also observe performance cuts and memory challenges that arise from storing and processing the additional paths' information. Other possible improvements to the quality of solutions could be made by examining the types of solutions returned by the algorithms when using different types of dissimilarity metrics and constraints. With several other types of dissimilarity constraints available, it is possible that these metrics may yield substantially different set of solutions.
		
		The A* component of the algorithms combined with both the Simple Height Restriction and the Expanding Height Restriction reduced the computational time of the algorithms. The results from the Expanding Height Restriction are not included in this paper as they were not competitive with the Simple Height Restriction. Both height restrictions were able to solve the same problems, but the additional overhead required for the Expanding Height Restriction meant that it took longer. In some cases the height restriction algorithms were able to find more solutions within the preset time-out time than their regular counterparts. Our aggressive choice of height restriction parameters for the Simple Height Restriction provided a significant improvement in time without decreasing the quality of solutions. Another appeal of the Simple Height Restriction is its simplicity in implementation.

		Before the Bidirectional Selection Method could be efficiently used in practice further refinements would be necessary.  It is worth noting that the largest test problem we considered was only 3200 meters.  The time to solve such problems was approximately several hours.   One possible direction for improvement would be to massively parallelize the algorithm to reduce the running time. 
		
		Improving the accuracy of the heuristic used for the A* component of the algorithm would decrease the overall running time. The heuristic could be improved by the addition of unavoidable cost factors such as rivers, or mountain chains separating the source and the destination. 
		
		Future versions of the dissimilar multipath algorithms could be modified to connect two sections of an existing road network. This process would be straight-forward assuming the cost of connecting the new road to a specific point of an existing network is known. Instead of simply adding one starting vertex, all candidate vertices along the existing road could be added with their respective costs. The destination nodes code similarly be modified to trigger the stopping condition. Existing road networks bisecting the search space could also be incorporating by modifying the terrain graph to include low-cost or no-cost edges, as appropriate.

		\section*{Acknowledgments}
		\label{sec:Acknowledgments}
		
			Numerical experiments were run on Westgrid (part of ComputeCanada) and in the CA$^2$ lab (funded through a Canadian Foundation for Innovation (CFI) Leaders Opportunity grant and a BC Knowledge Development grant). Pushak was supported by a Natural Science and Engineering Research Council (NSERC) of Canada Undergraduate Research Award (USRA), an Irving K. Barber School Undergraduate Research Award (URA) and an NSERC Discovery grant from Lucet (\#298145-2013). Lucet was partly supported by an NSERC Discovery grant (\#298145-2013). Hare was partly supported by an NSERC Discovery grant (\#355571-2013).
			
			The authors would like to thank Softree Technical Systems Inc. for numerous fruitful discussions on the road design problem as part of an ongoing NSERC Collaborative and Research Development grant (\#CRDPJ 411318-10) between Softree, Hare, and Lucet.  The authors would like to thank the anonymous referees for their time and valuable feedback.

\bibliographystyle{abbrv}
\bibliography{bibliography}

\begin{thebibliography}{10}

\bibitem{ABRAHAM-13}
I.~Abraham, D.~Delling, A.~V. Goldberg, and R.~F. Werneck.
\newblock Alternative routes in road networks.
\newblock {\em Journal of Experimental Algorithmics}, 18(1.3):1--17, 2013.

\bibitem{AKGUN-00}
V.~Akg\"un, E.~Erkut, and R.~Batta.
\newblock On finding dissimilar paths.
\newblock {\em European Journal of Operational Research}, 121(2):232--246,
  2000.

\bibitem{BOSURGI-13}
G.~Bosurgi, O.~Pellegrino, and G.~Sollazzo.
\newblock A {PSO} highway alignment optimization algorithm considering
  environmental constraints.
\newblock {\em Advances in Transportation Studies}, {}(31):63 -- 80, 2013.

\bibitem{BOSURGI-14}
G.~Bosurgi, O.~Pellegrino, and G.~Sollazzo.
\newblock Using genetic algorithms for optimizing the {PPC} in the highway
  horizontal alignment design.
\newblock {\em Journal of Computing in Civil Engineering},
  http://dx.doi.org/10.1061/(ASCE)CP.1943-5487.0000452, 2014.

\bibitem{BRANDER-95}
A.~W. Brander and M.~C. Sinclair.
\newblock A comparative study of $k$-shortest path algorithms.
\newblock In {\em Proc. of 11th UK Performance Engineering Workshop}, pages
  370--379, 1995.

\bibitem{CHURCH-92}
R.~L. Church, S.~R. Loban, and K.~Lombard.
\newblock An interface for exploring spatial alternatives for a corridor
  location problem.
\newblock {\em Computers \& Geosciences}, 18:1095--1105, 1992.

\bibitem{DELLING-09}
D.~Delling, P.~Sanders, D.~Schultes, and D.~Wagner.
\newblock Engineering route planning algorithms.
\newblock In J.~Lerner, D.~Wagner, and K.~Zweig, editors, {\em Algorithmics of
  Large and Complex Networks}, volume 5515 of {\em Lecture Notes in Computer
  Science}, pages 117--139. Springer Berlin Heidelberg, 2009.

\bibitem{DELL-05}
P.~Dell'Olmo, M.~Gentili, and A.~Scozzari.
\newblock On finding dissimilar {P}areto-optimal paths.
\newblock {\em European Journal of Operational Research}, 162:70--82, 2005.

\bibitem{DOLAN-02}
E.~D. Dolan and J.~J. Mor{\'e}.
\newblock Benchmarking optimization software with performance profiles.
\newblock {\em Mathematical Programming}, 91(2):201--213, 2002.

\bibitem{DUARTE-07}
A.~Duarte and R.~Marti.
\newblock Tabu search and {GRASP} for the maximum diversity problem.
\newblock {\em European Journal of Operational Research}, 178:71 -- 84, 2007.

\bibitem{EASA-08}
S.~M. Easa and A.~Mehmood.
\newblock Optimizing design of highway horizontal alignments: New substantive
  safety approach.
\newblock {\em Computer-Aided Civil and Infrastructure Engineering},
  23(7):560--573, 2008.

\bibitem{eppstein97}
D.~Eppstein.
\newblock Finding the $k$ shortest paths.
\newblock {\em SIAM Journal on Computing}, 28(2):652--673, 1998.

\bibitem{HARE-14}
W.~Hare, S.~Hossain, Y.~Lucet, and F.~Rahman.
\newblock Models and strategies for efficiently determining an optimal vertical
  alignment of roads.
\newblock {\em Computers \& Operations Research}, 44(0):161 -- 173, 2014.

\bibitem{HARE-14b}
W.~Hare, Y.~Lucet, and F.~Rahman.
\newblock A mixed-integer linear programming model to optimize the vertical
  alignment considering blocks and side-slopes in road construction.
\newblock Technical report, University of British Columbia, 2014.

\bibitem{HARE-11}
W.~L. Hare, V.~R. Koch, and Y.~Lucet.
\newblock Models and algorithms to improve earthwork operations in road design
  using mixed integer linear programming.
\newblock {\em European Journal of Operational Research}, 215(2):470 -- 480,
  2011.

\bibitem{HOFFMAN-59}
W.~Hoffman and R.~Pavley.
\newblock A method for the solution of the nth best path problem.
\newblock {\em Journal of the ACM}, 6:506--514, 1959.

\bibitem{HUBER-85}
D.~L. Huber and R.~L. Church.
\newblock Transmission corridor location modeling.
\newblock {\em Journal of Transportation Engineering}, 111:114--130, 1985.

\bibitem{ikeda94}
T.~Ikeda, M.-Y. Hsu, H.~Imai, S.~Nishimura, H.~Shimoura, T.~Hashimoto,
  K.~Tenmoku, and K.~Mitoh.
\newblock A fast algorithm for finding better routes by {AI} search techniques.
\newblock In {\em Vehicle navigation and information systems conference}, 1994.

\bibitem{JHA-03}
M.~K. Jha.
\newblock Criteria-based decision support system for selecting highway
  alignments.
\newblock {\em Journal of Transportation Engineering}, 129(1):33--41, 2003.

\bibitem{JHA-06}
M.~K. Jha and E.~Kim.
\newblock Highway route optimization based on accessibility, proximity, and
  land-use changes.
\newblock {\em Journal of Transportation Engineering}, 132(5):435--439, 2006.

\bibitem{JHA-04}
M.~K. Jha and P.~Schonfeld.
\newblock A highway alignment optimization model using geographic information
  systems.
\newblock {\em Transportation Research Part A}, 38(6):455--481, 2004.

\bibitem{JHA-06a}
M.~K. Jha, P.~Schonfeld, J.-C. Jong, and E.~Kim.
\newblock {\em Intelligent Road Design}, volume~19.
\newblock WIT Press, 2006.

\bibitem{JOHNSON-92}
P.~Johnson, D.~Joy, D.~Clarke, and J.~Jacobi.
\newblock {\em An Enhanced Highway Routing Model: Program, Description,
  Methodology, and Revised User's Manual}.
\newblock Oak Ridge National Laboratory, 1992.

\bibitem{JONG-98}
J.-C. Jong.
\newblock {\em Optimizing Highway Alignments with Genetic Algorithms}.
\newblock PhD thesis, University of Maryland, 1998.

\bibitem{JONG-03}
J.-C. Jong and P.~Schonfeld.
\newblock An evolutionary model for simultaneously optimizing three-dimensional
  highway alignments.
\newblock {\em Transportation Research Part B: Methodological}, 37(2):107 --
  128, 2003.

\bibitem{KANG-08}
M.~W. Kang.
\newblock {\em An alignment optimization model for a simple highway network}.
\newblock PhD thesis, University of Maryland, 2008.

\bibitem{KANG-12}
M.-W. Kang, M.~K. Jha, and P.~Schonfeld.
\newblock Applicability of highway alignment optimization models.
\newblock {\em Transportation Research Part C: Emerging Technologies},
  21(1):257 -- 286, 2012.

\bibitem{KANG-07}
M.~W. Kang, P.~Schonfeld, and J.-C. Jong.
\newblock Highway alignment optimization through feasible gates.
\newblock {\em Journal of Advanced Transportation}, 41(2):115--144, 2007.

\bibitem{KANG-10}
M.~W. Kang, N.~Yang, P.~Schonfeld, and M.~Jha.
\newblock Bilevel highway route optimization.
\newblock {\em Transportation Research Record}, 2197(2197):107--117, 2010.

\bibitem{KANG-13}
Y.~Kang, R.~Batta, and C.~Kwon.
\newblock Generalized route planning model for hazardous material
  transportation with var and equity considerations.
\newblock {\em Computers \& Operations Research}, 43:237--247, March 2014.

\bibitem{KUBY-97}
M.~J. Kuby, X.~Zhongyi, and X.~Xiaodong.
\newblock A minimax method for finding the k best ``differentiated'' paths.
\newblock {\em Geographical Analysis}, 29:298--313, 1997.

\bibitem{LAWLER-72}
E.~L. Lawler.
\newblock A procedure for computing the {K} best solutions to discrete
  optimization problems and its application to the shortest path problem.
\newblock {\em Management Science}, 18(7):401--405, 1972.

\bibitem{LEE-09}
Y.~Lee, Y.-R. Tsou, and H.-L. Liu.
\newblock Optimization method for highway horizontal alignment design.
\newblock {\em Journal of Transportation Engineering}, 135(4):217--224, 2009.

\bibitem{LOMBARD-93}
K.~Lombard and R.~L. Church.
\newblock The gateway shortest path problem: Generating alternative routes for
  a corridor location problem.
\newblock {\em Geographical Systems}, 1:22--45, 1993.

\bibitem{MARTI-14}
R.~Marti, V.~Campos, M.~G. Resende, and A.~Duarte.
\newblock Multiobjective {GRASP} with path relinking.
\newblock {\em European Journal of Operational Research}, 240(1):54--71, 2015.

\bibitem{MARTI-09}
R.~Marti, J.~L.~G. Velarde, and A.~Duarte.
\newblock Heuristics for the bi-objective path dissimilarity problem.
\newblock {\em Computers \& Operations Research}, 36:2905--2912, 2009.

\bibitem{MONDAL-14}
S.~Mondal, Y.~Lucet, and W.~Hare.
\newblock Optimizing horizontal alignment of roads in a specified corridor.
\newblock Technical report, University of British Columbia, 2014.

\bibitem{ROUPHAIL-96}
N.~Rouphail, S.~Ranjithan, W.~El~Dessouki, and T.~Smight.
\newblock A decision support system for dynamic pre-trip route planning.
\newblock In {\em Applications of Advanced Technologies in Transportation
  Engineering - International Conference}, 1996.

\bibitem{SCAPARRA-14}
M.~P. Scaparra, R.~L. Church, and F.~A. Medrano.
\newblock Corridor location: the multi-gateway shortest path model.
\newblock {\em Journal of Geographical Systems}, 16:287--309, 2014.

\bibitem{SHAFAHI-13}
Y.~Shafahi and M.~Bagherian.
\newblock A customized particle swarm method to solve highway alignment
  optimization problem.
\newblock {\em Computer-Aided Civil \& Infrastructure Engineering}, 28(1):52 --
  67, 2013.

\bibitem{TRIETSCH-87a}
D.~Trietsch.
\newblock Comprehensive design of highway networks.
\newblock {\em Transportation Science}, 21(1):26--35, 1987.

\bibitem{TRIETSCH-87}
D.~Trietsch.
\newblock A family of methods for preliminary highway alignment.
\newblock {\em Transportation Science}, 21(1):17--25, 1987.

\bibitem{TURNER-78}
A.~K. Turner.
\newblock A decade of experience in computer aided route selection.
\newblock {\em Photogrammetric Engineering And Remote Sensing},
  44(12):1561--1576, 1978.

\bibitem{nationalMap}
USGS.
\newblock The national map viewer.
\newblock http://viewer.nationalmap.gov.
\newblock Accessed: 2015-05-26.

\bibitem{westgrid}
Westgrid.
\newblock http://www.westgrid.ca.
\newblock Accessed: 2014-09-13.

\bibitem{ksp}
Wikipedia.
\newblock http://en.wikipedia.org/wiki/K\_shortest\_path\_routing.
\newblock Accessed: 2014-09-13.

\bibitem{YANG-14}
N.~Yang, M.-W. Kang, P.~Schonfeld, and M.~K. Jha.
\newblock Multi-objective highway alignment optimization incorporating
  preference information.
\newblock {\em Transportation Research Part C: Emerging Technologies},
  40(0):36--48, 2014.

\bibitem{YEN-71}
J.~Y. Yen.
\newblock Finding the {K} shortest loopless paths in a network.
\newblock {\em Management Science}, 17(11):712--716, 1971.

\bibitem{ZHANG-05}
X.~Zhang and M.~P. Armstrong.
\newblock Using a genetic algorithm to generate alternatives for
  multi-objective corridor location problems.
\newblock In {\em Proceedings of the 8th International Conference on
  GeoComputation}, 2005.

\end{thebibliography}

\newpage
\appendix
\appendixpage
\addappheadtotoc
	\section{Test Sets}
	\label{app:Test Sets}
		
		We have defined three terrain types to categorize the test maps to ensure that we have a diverse set of tests. With the interests of road construction in mind we have based this upon our maximum road grade of 10\%.
		Let $\mxy$ be the maximum grade of the terrain in each of the eight $\h$ directions allowed for edges at the point $(\x,\y)$.
		Then for each position $(\x_i,\y_i)$ corresponding to the location of a vertex in our map, we calculate $\mxiyi$.
		\begin{df}
		Let $\T(\x_i,\y_i)$ be the type of terrain at position $(\x_i,\y_i)$.
		\begin{equation}
		\label{eq:terrain}
		\T(\x_i,\y_i) =
		\begin{cases}
		\text{A,} & \text{if } \mxiyi \leq 10 \%, \\
		\text{B,}  & \text{if } 10 \% <  \mxiyi < 20 \%, \\
		\text{C,} & \text{if }  20 \% \leq  \mxiyi. \\
		\end{cases}
		\end{equation}
		\end{df}	
		
		Using these definitions we then calculated the percentage of each category present in each test map.
		
		\subsection{Test Set One}
		\label{app:Test Set One}
			\begin{table}[H]
				\centering
				\begin{tabular}{|r|rrrrrr|}
					\hline
					& Dim x & Dim y & Dim z & A     & B     & C \\
					\hline
					Test 1 & 40    & 20    & 5     & 100\% & 0\%   & 0\% \\
					Test 2 & 40    & 5     & 9     & 93\%  & 7\%   & 0\% \\
					Test 3 & 40    & 20    & 23    & 54\%  & 27\%  & 19\% \\
					Test 4 & 40    & 20    & 37    & 12\%  & 56\%  & 32\% \\
					Test 5 & 40    & 5     & 38    & 0\%   & 0\%   & 100\% \\
					Test 6 & 40    & 20    & 53    & 6\%   & 43\%  & 52\% \\
					Test 7 & 40    & 20    & 60    & 0\%   & 8\%   & 92\% \\
					Test 8 & 40    & 5     & 60    & 15\%  & 15\%  & 70\% \\
					Test 9 & 40    & 20    & 88    & 0\%   & 20\%  & 80\% \\
					Test 10 & 40    & 20    & 90    & 0\%   & 1\%   & 99\% \\
					Test 11 & 80    & 40    & 11    & 91\%  & 9\%   & 0\% \\
					Test 12 & 80    & 10    & 10    & 92\%  & 6\%   & 2\% \\
					Test 13 & 80    & 40    & 28    & 37\%  & 31\%  & 32\% \\
					Test 14 & 80    & 40    & 39    & 46\%  & 43\%  & 11\% \\
					Test 15 & 80    & 10    & 41    & 29\%  & 56\%  & 16\% \\
					Test 16 & 80    & 40    & 67    & 5\%   & 44\%  & 52\% \\
					Test 17 & 80    & 40    & 84    & 38\%  & 16\%  & 47\% \\
					Test 18 & 80    & 10    & 61    & 18\%  & 22\%  & 60\% \\
					Test 19 & 80    & 40    & 101   & 17\%  & 21\%  & 61\% \\
					Test 20 & 80    & 40    & 123   & 54\%  & 4\%   & 43\% \\
					Test 21 & 160   & 80    & 5     & 100\% & 0\%   & 0\% \\
					Test 22 & 160   & 20    & 14    & 100\% & 0\%   & 0\% \\
					Test 23 & 160   & 80    & 36    & 71\%  & 19\%  & 10\% \\
					Test 24 & 160   & 80    & 50    & 69\%  & 22\%  & 9\% \\
					Test 25 & 160   & 20    & 31    & 69\%  & 28\%  & 3\% \\
					Test 26 & 160   & 80    & 85    & 49\%  & 30\%  & 21\% \\
					Test 27 & 160   & 80    & 97    & 41\%  & 30\%  & 30\% \\
					Test 28 & 160   & 20    & 27    & 87\%  & 13\%  & 0\% \\
					Test 29 & 160   & 80    & 127   & 46\%  & 19\%  & 35\% \\
					Test 30 & 160   & 80    & 146   & 16\%  & 37\%  & 46\% \\
					\hline
				\end{tabular}%
				\caption{The dimensions and type of terrain for the thirty tests used in the first test set.}
				\label{tab:Test Set 1}%
			\end{table}%

		\subsection{Test Set Two}
		\label{app:Test Set Two}
			\begin{table}[H]
				\centering
				\begin{tabular}{|r|rrrrrr|}
					\hline
					& Dim x & Dim y & Dim z & A     & B     & C \\
					\hline
					Test 1 & 320   & 160   & 9     & 99\%  & 1\%   & 0\% \\
					Test 2 & 320   & 40    & 20    & 100\% & 0\%   & 0\% \\
					Test 3 & 320   & 160   & 41    & 75\%  & 21\%  & 4\% \\
					Test 4 & 320   & 160   & 67    & 67\%  & 20\%  & 12\% \\
					Test 5 & 320   & 40    & 47    & 78\%  & 10\%  & 12\% \\
					Test 6 & 320   & 160   & 111   & 49\%  & 24\%  & 27\% \\
					Test 7 & 320   & 160   & 122   & 39\%  & 34\%  & 28\% \\
					Test 8 & 320   & 40    & 132   & 55\%  & 17\%  & 28\% \\
					Test 9 & 320   & 160   & 165   & 39\%  & 24\%  & 37\% \\
					Test 10 & 320   & 160   & 195   & 41\%  & 23\%  & 36\% \\
					Test 11 & 640   & 320   & 49    & 71\%  & 22\%  & 6\% \\
					Test 12 & 640   & 80    & 44    & 94\%  & 5\%   & 1\% \\
					Test 13 & 640   & 320   & 107   & 68\%  & 18\%  & 14\% \\
					Test 14 & 640   & 320   & 192   & 63\%  & 19\%  & 18\% \\
					Test 15 & 640   & 80    & 188   & 64\%  & 24\%  & 12\% \\
					Test 16 & 640   & 320   & 274   & 32\%  & 21\%  & 47\% \\
					Test 17 & 640   & 320   & 319   & 20\%  & 27\%  & 53\% \\
					Test 18 & 640   & 80    & 332   & 6\%   & 27\%  & 66\% \\
					Test 19 & 640   & 320   & 416   & 7\%   & 17\%  & 75\% \\
					Test 20 & 640   & 320   & 479   & 18\%  & 29\%  & 53\% \\
					\hline
				\end{tabular}%
				\caption{The dimensions and type of terrain for the twenty tests used in the second test set.}
				\label{tab:Test Set 2}%
			\end{table}%

\end{document}